\documentclass[twocolumn]{aastex631}

\shorttitle{X-ray absorption in GRBs}
\shortauthors{Valan, Larsson \& Ahlgren}
\graphicspath{{./}{figures/}}

\begin{document}

\title{Investigating time variability of X-ray absorption in \textit{Swift} GRBs}

\author[0000-0002-7653-9235]{Vlasta Valan}
\affiliation{Department of Physics, KTH Royal Institute of Technology, The Oskar Klein Centre, AlbaNova, SE-106 91 Stockholm, Sweden}

\author[0000-0003-0065-2933]{Josefin Larsson}
\affiliation{Department of Physics, KTH Royal Institute of Technology, The Oskar Klein Centre, AlbaNova, SE-106 91 Stockholm, Sweden}

\author[0000-0003-4000-8341]{Bj\"{o}rn  Ahlgren}
\affiliation{Department of Physics, KTH Royal Institute of Technology, The Oskar Klein Centre, AlbaNova, SE-106 91 Stockholm, Sweden}



\begin{abstract}

The existence of excess absorption in the X-ray spectra of GRBs is well known, but the primary location of the absorbing material is still uncertain.  To gain more knowledge about this, we have performed a time-resolved analysis of the X-ray spectra of 199 GRBs observed by the \textit{Swift} X-ray telescope, searching for evidence of a decreasing column density ($N_{\mathrm{H,intr}}$) that would indicate that the GRBs are ionizing matter in their surroundings. 
We structured the analysis as Bayesian inference and used an absorbed power-law as our baseline model. We also explored alternative spectral models in cases where decreasing absorption was inferred.
The analysis reveals seven GRBs that show signs of a decrease in $N_{\mathrm{H,intr}}$, but we note that alternative models for the spectral evolution cannot be ruled out. We conclude that the excess absorption in the vast majority of GRBs must originate on large scales of the host galaxies and/or in the intergalactic medium. Our results also imply that an evolving column density is unlikely to affect the spectral analysis of the early X-ray spectra of GRBs. In line with this, we show that estimating the total $N_{\mathrm{H,intr}}$ from early {\it Swift} data in Window Timing mode reveals the same increasing trend with redshift as previous results based on data taken at later times, but with tighter constraints. 

\end{abstract}


\section{Introduction} \label{sec:intro}

Gamma-ray bursts (GRBs) are produced in connection with supernova explosions and compact object mergers \citep{Galama1998, Bloom2002, Abbott2017}. Their emission is divided into two distinct phases: prompt emission and an afterglow. The prompt emission arises from a relativistic jet, has a short duration of seconds to minutes and is primarily observed in gamma rays \cite[e.g.,][]{Kumar2015}. The afterglow arises when the relativistic jet interacts with the surrounding medium, which produces synchrotron radiation that is observed across all wavelengths from X-rays to radio on a longer time scale \citep{Sari1998}.

The X-ray emission from GRBs is regularly monitored by the \textit{Neils Gehrles Swift} satellite and its X-ray telescope (XRT, \citealt{Gehrles2004, Burrows2005}). The XRT observations start $\sim 100$~s after the trigger and cover the $0.3-10$~keV energy range. GRB spectra in this energy range are mostly well described by an absorbed power-law model \citep{Evans2009, Racusin2009}, where the absorption is usually expressed as a combination of Galactic \citep{Kalberla2005, Willingale2013} and extragalactic (intrinsic) components.  The intrinsic absorption was first recorded by \textit{BeppoSax} as an excess above the Galactic contribution \citep{Frontera2000,Stratta2004}. One of the key results obtained from \textit{Swift} is that intrinsic X-ray absorption is ubiquitous in GRBs and that it increases with redshift \citep{Campana2010, Watson2012, Starling2013, Rahin2019, Dalton2020}. This absorption has attracted considerable interest as it can reveal information about the GRB environments and host galaxies, as well as intervening matter between the Milky Way and the hosts. 

The absorption leads to a reduction of photons at low energies, typically $<2$ keV, but individual absorption lines are not detected with current instruments \citep{Campana2016}. The amount of intrinsic absorption is commonly parameterized in terms of $N_{\mathrm{H,intr}}$, which is the hydrogen column density along the line-of-sight (LOS) outside our Galaxy. However, it should be noted that the absorption cross section is dominated by metals. The detailed properties of the absorbing material cannot be constrained from typical XRT spectra, which means that $N_{\mathrm{H,intr}}$ is usually determined based on a number of simplifying assumptions. 
In particular, it is common practice to assume solar metallicity for the host galaxies of GRBs, with abundances taken from \citet{Wilms2000}. This assumption means that the value of $N_{\mathrm{H,intr}}$ is underestimated as the host galaxies of GRBs usually have sub-solar metallicities \citep{Starling2013, Tanga2016, Bignone2017, Nugent2022}. Another simplifying assumption is that the absorbing gas is neutral  \citep{Behar2011, Schady2011, Starling2013}. Ionized gas has a lower cross-section for X-ray photons and would thus need a larger column density to produce the same opacity. Assuming neutral gas thus results in the total $N_{\mathrm{H,intr}}$ being underestimated \citep{Schady2011}.

While the presence of excess absorption and its correlation with redshift in GRB spectra is well established, its primary origin is still a source of debate. Some authors argue that the GRB hosts can account for the absorption and redshift evolution \citep{Schady2011, Watson2012, Watson2013}. Others argue that a significant part of the absorption is due to the full integrated LOS, which includes diffuse inter-galactic medium (IGM) and intervening objects \citep{Behar2011, Campana2012, Starling2013, Campana2015, Dalton2021}. 

One way to probe the location of the absorbing medium is to use time-resolved measurements of the column density. Any material located in the vicinity of the progenitor that is not already ionized at the time of the GRB is expected be ionized by the X-ray/UV photons in the prompt emission and afterglow. Consequently, the medium into which the afterglow emission propagates at later times is more ionized than at earlier times, leading to a decrease in absorption with time \citep{Perna1998, Lazatti2002, Perna2002}. This in turn can help us understand the progenitor environment: a more compact absorbing region leads to a faster decrease in $N_{\mathrm{H,intr}}$, which can be used to infer the density of the medium when combined with the values of $N_{\mathrm{H,intr}}$. Evidence of a decreasing $N_{\mathrm{H,intr}}$ has been reported for a small number of GRBs \citep{Starling2005, Grupe2007, Campana2007, Campana2021}. However, the early results were questioned by \citet{Butler2007}, who showed that intrinsic spectral evolution may be misinterpreted as a decreasing $N_{\mathrm{H,intr}}$ and that performing the analysis on a finer time scale reveals unphysical variations in $N_{\mathrm{H,intr}}$. 

In this paper we present the first systematic study of time variability of $N_{\mathrm{H,intr}}$ in the X-ray spectra of a large sample of GRBs. We perform a finely time-resolved spectral analysis of 199  GRBs observed by \textit{Swift}. We use the Bayesian inference method, which allows us to investigate the posterior distribution of $N_{\mathrm{H,intr}}$ and possible degeneracies between spectral parameters. The analysis focuses on early \textit{Swift}~XRT data in the Window Timing (WT) mode, which probes time scales down to $\sim 20$ s after the trigger in the rest frame of the GRBs. It is important to note that our reported values $N_{\mathrm{H,intr}}$ should be considered as lower limits due to the assumptions mentioned above. However, this does not affect the results regarding the evolution of $N_{\mathrm{H,intr}}$ with time. 

The paper is organized as follows: we present the sample in Section \ref{sec:sample}, followed by a description of the analysis methods in Section \ref{sec:analysis}. Our results are summarized in Section \ref{sec:results}. We discuss the interpretation of our results in Section \ref{sec:discussion} and finalize the paper with conclusions in Section \ref{sec:conclusions}.

\section{Sample selection and data reduction} \label{sec:sample}

The sample comprises all GRBs analyzed in \cite{Valan2018} (V18) and \cite{Valan2021} (V21). It consists of 199 GRBs observed between 2005 Apr 01 and 2018 Dec 31. All the GRBs have spectroscopic redshifts and XRT WT average fluxes higher than $2 \times 10^{-10} \ \mathrm{erg \ cm^{-2} \ s^{-1}}$. As a result of these selection criteria, the sample is primarily composed of long GRBs (196 are long and 3 are short). Data were downloaded from the \textit{Swift} UK Science Data Centre repository\footnote{\url{https://www.swift.ac.uk/xrt_spectra/}} and reduced using the automatic pipeline \citep{Evans2009}. In order to perform time-resolved spectral analysis, time intervals with approximately constant count rates were identified using the Bayesian blocks algorithm \citep{Scargle1998}. The known potential calibration issues for XRT data at low energies ($< 0.6$ keV) were investigated in detail, as described in V18 and V21. The reader is also referred to those papers for further details about the data processing. The only difference in this work is that the spectra were grouped to have at least 1 count per bin in order to use the cstat fit statistic.

\section{Data Analysis} \label{sec:analysis}

We performed a time-resolved spectral analysis of the XRT WT data.  All GRBs in the sample have at least 3 time bins, which allows us to probe the time evolution of $N_{\mathrm{H,intr}}$. We fitted the XRT spectra in the energy range 0.3--10~keV, though the upper energy boundary was lowered if the signal stopped before 10~keV. Additionally, the lower boundary was set to 0.6~keV in three GRBs (GRB~151027A, GRB~170519A and GRB~180329B) due to calibration issues (see V21). 

The analysis was set up as a Bayesian inference procedure. This means that it relies on Bayes theorem, which states that the posterior probability is
\begin{equation}
    \mathrm{Pr}(\Theta|y)  = \frac{\mathrm{Pr}(\Theta)\mathrm{Pr}(y|\Theta)}{\int \mathrm{Pr}(\Theta)\mathrm{Pr}(y|\Theta) d\Theta} \propto \mathrm{Pr}(\Theta)\mathrm{Pr}(y|\Theta)
\end{equation}
where $\Theta$ are the model parameters, $y$ the observed data, $\mathrm{Pr}(\Theta)$ the prior, $\mathrm{Pr}(y|\Theta)$ the likelihood (based on cstat in \textsc{xspec}), and the denominator is the marginalized likelihood, also referred to as evidence, $p(d|\mathcal{M}$), where $d$ are data and $\mathcal{M}$ is the model. In our analysis, we have used pyMultiNest \citep{Buchner2014}, a python implementation of MultiNest \citep{Feroz2008, Feroz2009} to sample from the model posterior using 600 live points. The number of live points was chosen after testing to ensure fit stability \citep{Ahlgren2019}. The analysis was performed using PyXspec, a python implementation of \textsc{heasarc xspec} 12.8.1g \citep{xspec}.  

Our baseline model for the fits is an absorbed power law. 
We assume that the absorption occurs in the Milky Way and in the host galaxy of the GRB. We thus include two absorption components in the model: a Galactic component with column density $N_{\mathrm{H,Gal}}$, and an intrinsic component with column density $N_{\mathrm{H,intr}}$. 

The Galactic component is accounted for by the \textsc{xspec} model \textit{tbabs} \citep{Wilms2000} with  $N_{\mathrm{H,Gal}}$ fixed to the value obtained using the  \textit{$N_{\rm{H,tot}}$} tool,\footnote{\url{http://www.swift.ac.uk/analysis/nhtot/index.php}} which includes contributions from both atomic and molecular H \citep{Willingale2013}. The \textit{tbabs} model assumes that the molecular component comprises $20\%$ of the total column density. In a small number of GRBs, where the the molecular component was $<10\%$ or $>30\%$, we have instead used the \textit{tbvarabs} model to set the atomic and molecular components separately. 

The intrinsic absorption is modelled using the \textit{ztbabs} model, with the redshift fixed at the redshift of the GRB and $N_{\mathrm{H,intr}}$  left free to vary. This model assumes a neutral absorber with solar abundances, which is unlikely to be valid for most GRBs. 
As noted in Section~\ref{sec:intro}, these assumptions imply that the values of $N_{\mathrm{H,intr}}$ are effectively lower limits. If the ionisation increases with time, the absorption affecting the spectra will decrease, which is the effect we are looking for. 

We have chosen our priors to be uninformative. For the photon index ($\Gamma$) we choose a uniform prior and for $N_{\rm{H,intr}}$ a log-uniform prior, i.e. 
\\ \begin{center}
    $\mathrm{Pr}(\Gamma) = U(0.5, 5)$\\
    $\mathrm{Pr}(\mathrm{log} N_\mathrm{H,intr, 22}) = U(0.001, 100)$.
\end{center}
\citet{Racusin2009} showed that the X-ray spectra of GRBs are well described with a power-law model with a typical photon index $1.5 \le \Gamma \le 3$. Our wider range for the prior on this parameter was set to also include the lower and higher values of $\Gamma$ found in a small number of GRBs, e.g. V18, V21. For $N_{\mathrm{H,intr}}$, we have chosen $10^{19}\ \mathrm{cm^{-2}}$ as the lower limit and $10^{24}\ \mathrm{cm^{-2}}$ as the upper limit, considering the range of values quoted in the literature \citep{Starling2013, Valan2018, Rahin2019, Tanvir2019,Valan2021}.

For GRBs that showed a decreasing $N_{\mathrm{H,intr}}$ when fitted with this model, we also performed fits with two alternative models with fixed $N_{\mathrm{H,intr}}$ (referred to as $N_{\mathrm{H,intr,fixed}}$ from here on). The values of $N_{\mathrm{H,intr,fixed}}$ were taken from V18 and V21, where it was determined by simultaneously fitting all the time-resolved spectra with $N_{\mathrm{H,intr}}$ tied, but the power-law parameters free to vary. Using this $N_{\mathrm{H,intr,fixed}}$, we first fitted an absorbed power law with the same priors on $\Gamma$ as before. The results of these fits offer a way to assess the significance of the decreasing $N_{\mathrm{H,intr}}$. Secondly, a cutoff power law with  $N_{\mathrm{H,intr,fixed}}$ was fitted to assess whether the apparent decrease of $N_{\mathrm{H,intr}}$ can be explained by variability of a more complex intrinsic spectrum. 
In this case we have  kept the uniform prior for $\Gamma$ and chosen a uniform prior for the cutoff energy ($E_{\mathrm{cut}}$):
\\ \begin{center}
    $\mathrm{Pr}(E_{\mathrm{cut}}) = U(0.1, 30)$\\
    $\mathrm{Pr}(\Gamma) = U(0.5, 5)$.
\end{center}
We have chosen the upper limit on the cut-off energy at 30~keV as higher values would not induce any curvature in the fitted energy range below 10~keV.
 
We use Bayes factor (BF) for comparison between the power-law models with free and fixed absorption. 
Bayes factor is defined as 

\begin{equation}
  \mathrm{log(BF_{01})} = log(p(d|\mathcal{M}_0)) - log(p(d|\mathcal{M}_1)), \label{eq:BF}
\end{equation}
for comparison between models 0 and 1. 
Our interpretation of Bayes factor and its value is based on the Jeffreys scale \citep{Jeffreys1961}. In short: if the value of $\mathrm{log(BF_{01})}$ is $< 0$ then model 1 is favored, and if it is $> 0$ model 0 is favored. We will consider it strong evidence for one or the other models if the value of $\mathrm{log(BF)} < -1$ or $\mathrm{log(BF)} > 1$. 

In cases where we have fitted the cutoff power-law model, the Bayes factor is not suitable for model comparison as it is sensitive to the priors, and the models involved probe different parameter spaces. 
The same is true in cases where we wish to compare with previous fits to a power law + blackbody from V18,V21.
Instead, we used the Akaike Information Criterion (AIC), which is non-sensitive to our choice of priors. The AIC is defined as: 
\begin{equation} 
\mathrm{AIC} = 2k - 2 \mathrm{ln}(L_{\mathrm{max}}) \label{eq:AIC1},
\end{equation} where $k$ is the number of free model parameters and $L_{\mathrm{max}}$ is the maximum likelihood. 
We compute the difference in AIC between models 0 and 1 as $\Delta \mathrm{AIC= AIC_{0} - AIC_{1}}$, where negative (positive) values implies preference for model 0 (1). We take $|\Delta \mathrm{AIC}|>20$ as a limit to indicate strong preference for either model, motivated by a comparison of fit residuals for different values of  $\Delta \mathrm{AIC}$.

In the following, all times are quoted in the rest-frame of the GRB, $t_{\mathrm{rest}} = t_{\mathrm{obs}}(1+z)^{-1}$. We take the posterior mean to represent the best-fit model. All reported uncertainties for point estimates correspond to the $1\sigma$ credible intervals around the mean unless stated otherwise. We have assumed a flat Universe with $H_0 = 67.3 \ \mathrm{km \ s^{-1} \ Mpc^{-1}}$, $\Omega_M = 0.315$, $\Omega_\Lambda = 0.685$ \citep{Planck2014}.

\section{Results} \label{sec:results}

\begin{figure*}[htb!]
\gridline{\fig{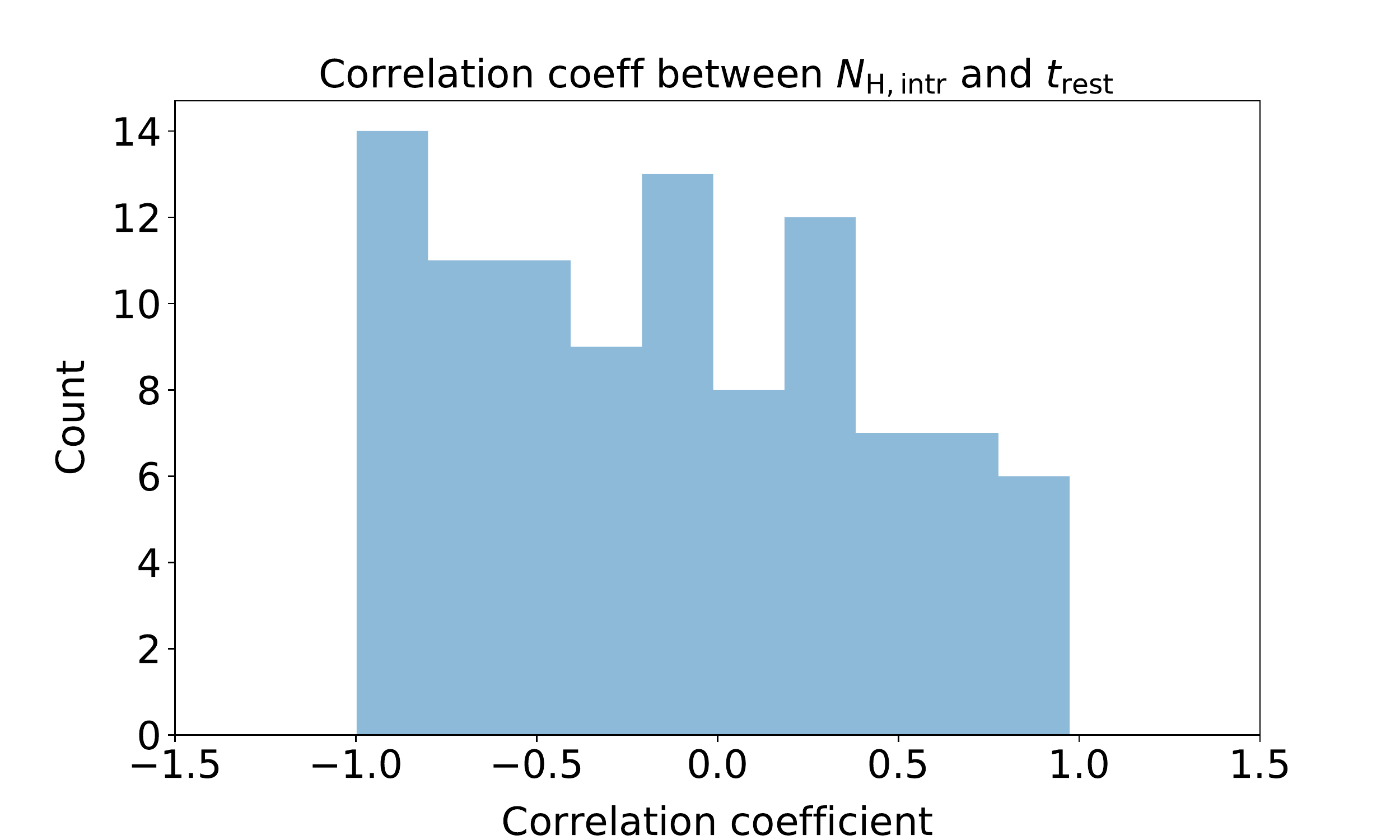}{0.49\textwidth}{}
          \fig{figures/Hist_corr_NH_Gamma.pdf}{0.49\textwidth}{}}
\caption{Histograms of Pearson correlation coefficients between $N_{\mathrm{H,intr}}$ and $t_{\rm rest}$ (left) and  $N_{\mathrm{H,intr}}$ and $\Gamma$ (right), obtained from the power-law fits with free $N_{\mathrm{H,intr}}$ for the full sample of GRBs. \label{fig:Histograms}}
\end{figure*}

As a first step in determining if $N_{\mathrm{H,intr}}$ decreases with time, we investigated their possible correlation using the Pearson correlation coefficient, the results of which are shown in Figure~\ref{fig:Histograms}. From the histogram it is clear that we observe a range of correlation coefficients between -1 and 1, with a slight preference toward the anticorrelated end (negative values). In Figure~\ref{fig:Histograms}, we also show the histogram of correlation coefficients between $N_{\mathrm{H,intr}}$ and $\Gamma$, showing that these two parameters have a tendency to be correlated. This is likely due to spectral degeneracies, driven by the fact that a highly absorbed soft spectrum can produce a similar count rate at low energies as a less absorbed hard spectrum. The degeneracies are discussed further below.  

We manually inspected all GRBs that had correlation coefficients between $N_{\mathrm{H,intr}}$ and time lower than -0.5, i.e. that showed a strong anticorrelation (28 GRBs in total). In order to select GRBs with credible declining $N_{\mathrm{H,intr}}$ in this step, we required a systematic decreasing trend in $N_{\mathrm{H,intr}}$ with time, as well as a total decrease of $N_{\mathrm{H,intr}}$ that was significant considering the $1 \sigma$ uncertainties. These criteria were fulfilled by nine GRBs. The main reason why the majority of GRBs with highly negative Pearson correlation coefficients were discarded is that these correlation coefficients do not account for parameter uncertainties, and the decrease was often consistent within error bars. 

\begin{figure*}[htb!]
\gridline{\fig{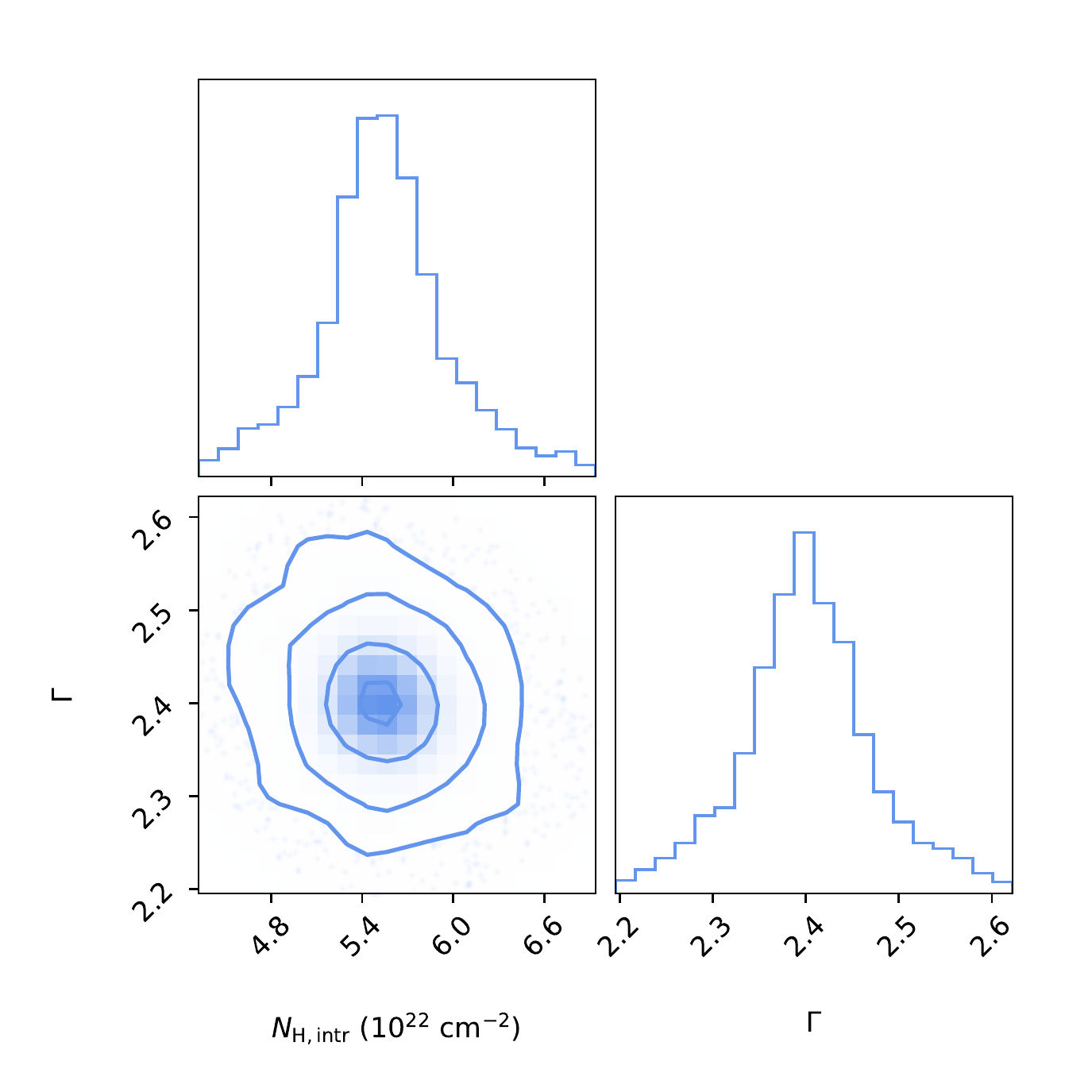}{0.49\textwidth}{}
            \fig{figures/GRB070306_bin2_fit1.pdf}{0.5\textwidth}{}}
\gridline{\fig{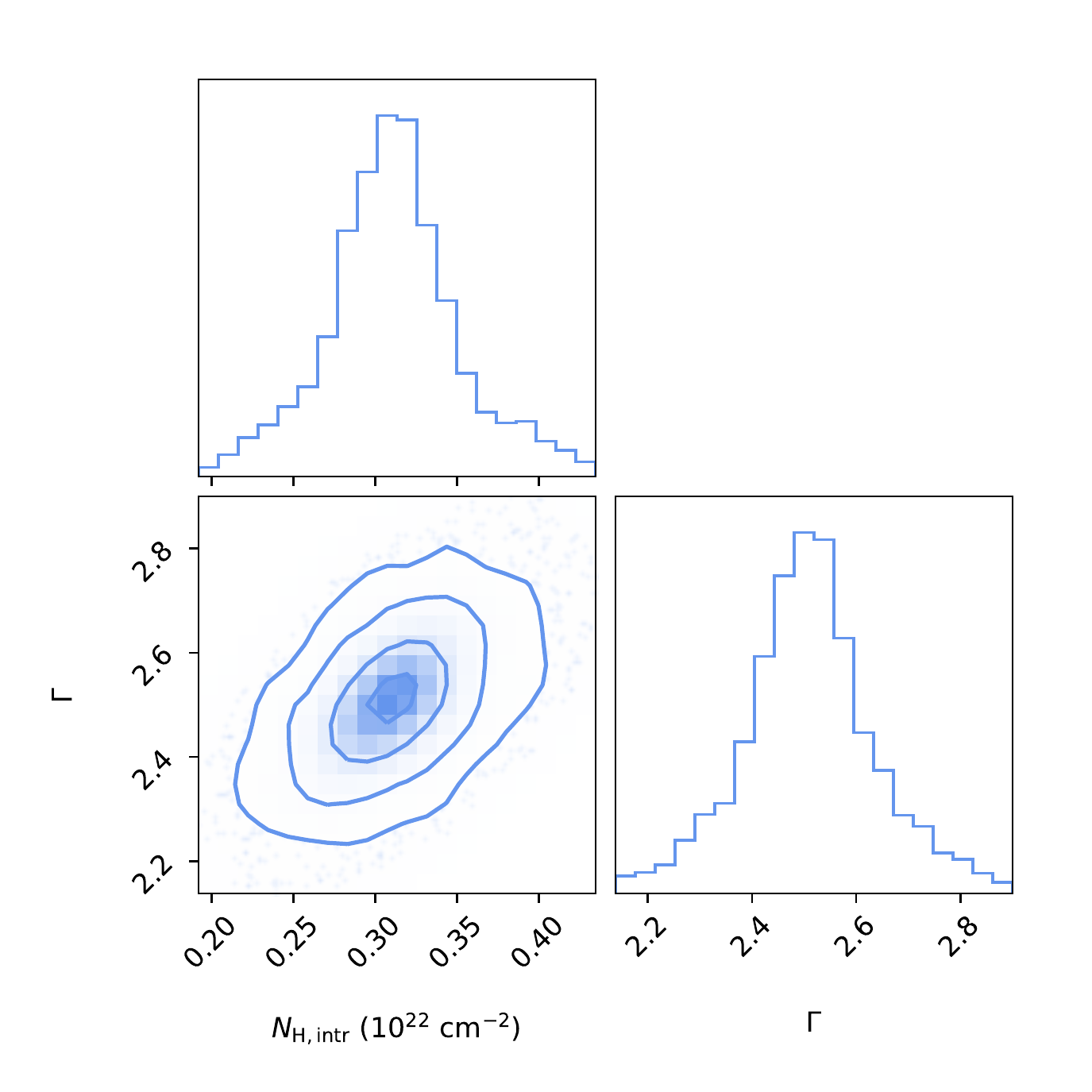}{0.49\textwidth}{}
    \fig{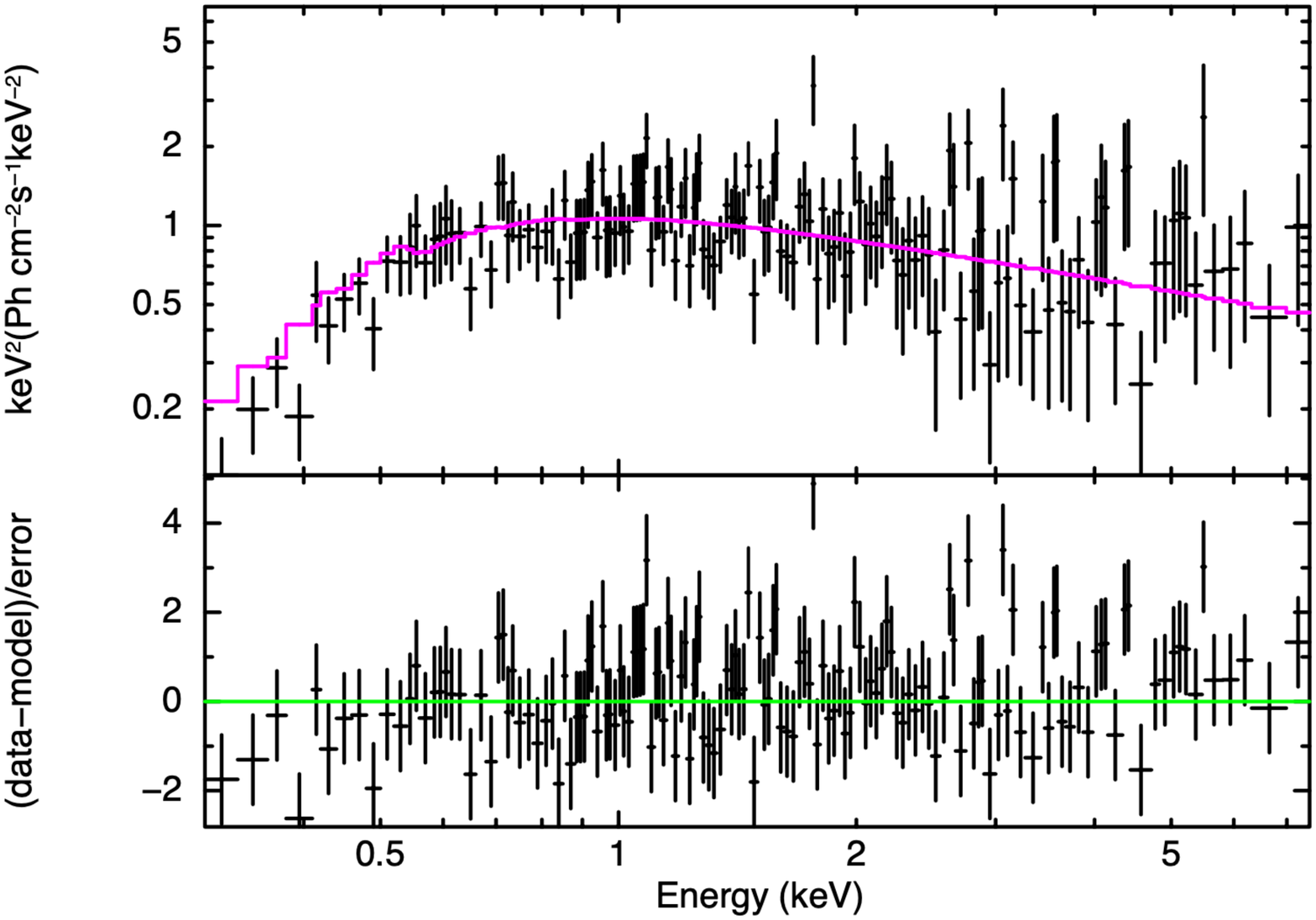}{0.5\textwidth}{}} 
\caption{Corner plots showing the posteriors for $N_{\rm{H,intr}}$ and $\Gamma$ (left) and the corresponding spectra with the best-fit model (right). GRB~070306 (top row, bin 2, time interval 67.1--72.5~s) illustrates the case where $N_{\rm{H,intr}}$ and $\Gamma$ are not degenerate, while GRB~170607A (bottom row, bin1, time interval 54.7--66.8~s) represents the case where degeneracies are present. Note that plotting the best-fit model in $\nu F_{\mathrm{\nu}}$ is not necessarily the best representation of the data and posterior, but is chosen as an illustration. 
\\ All corner plots and fits for all GRBs in Table~\ref{tab:finalsample} are available as online material. \label{fig:degeneracies}}
\end{figure*}

\begin{figure*}[htb!]
\gridline{\fig{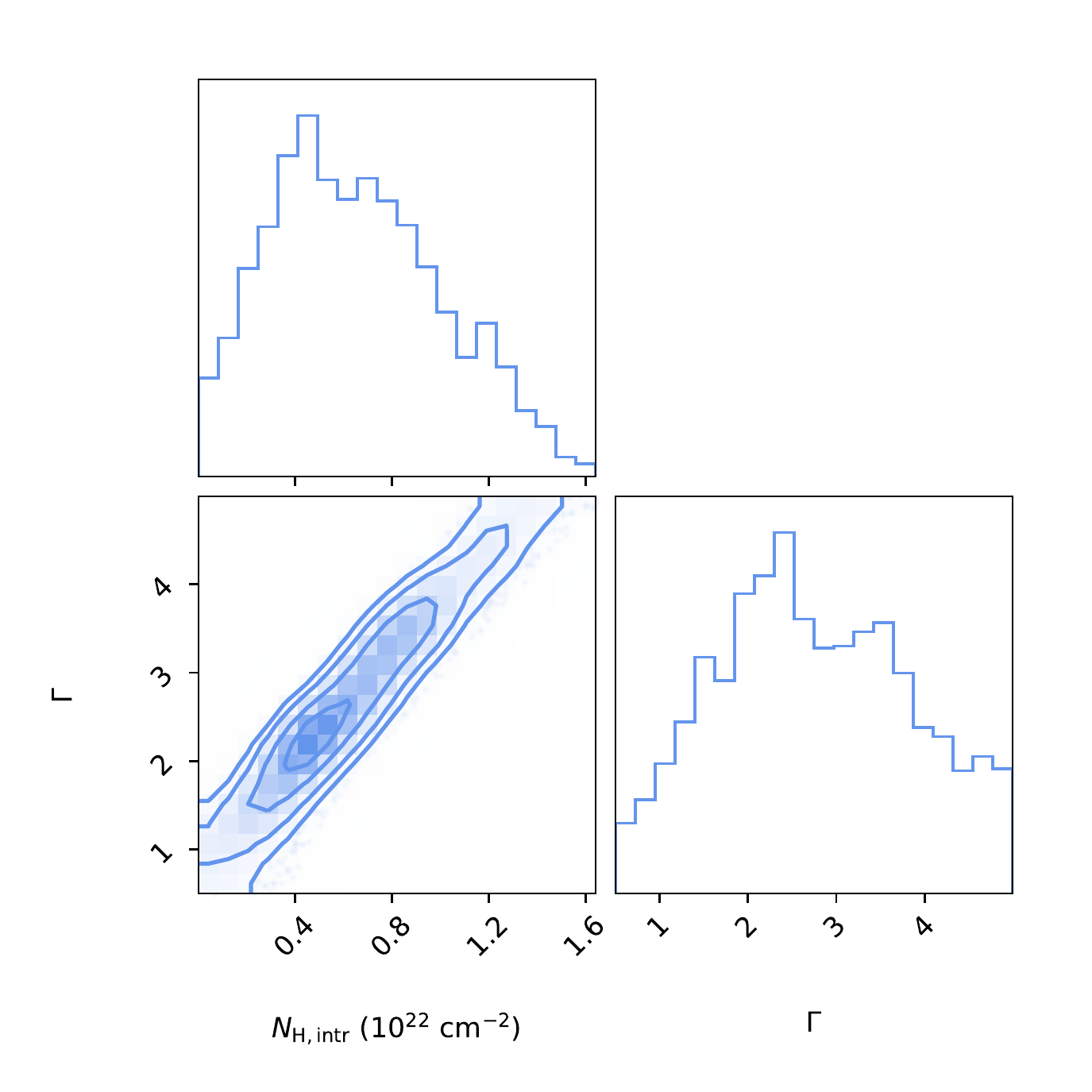}{0.49\textwidth}{}
    \fig{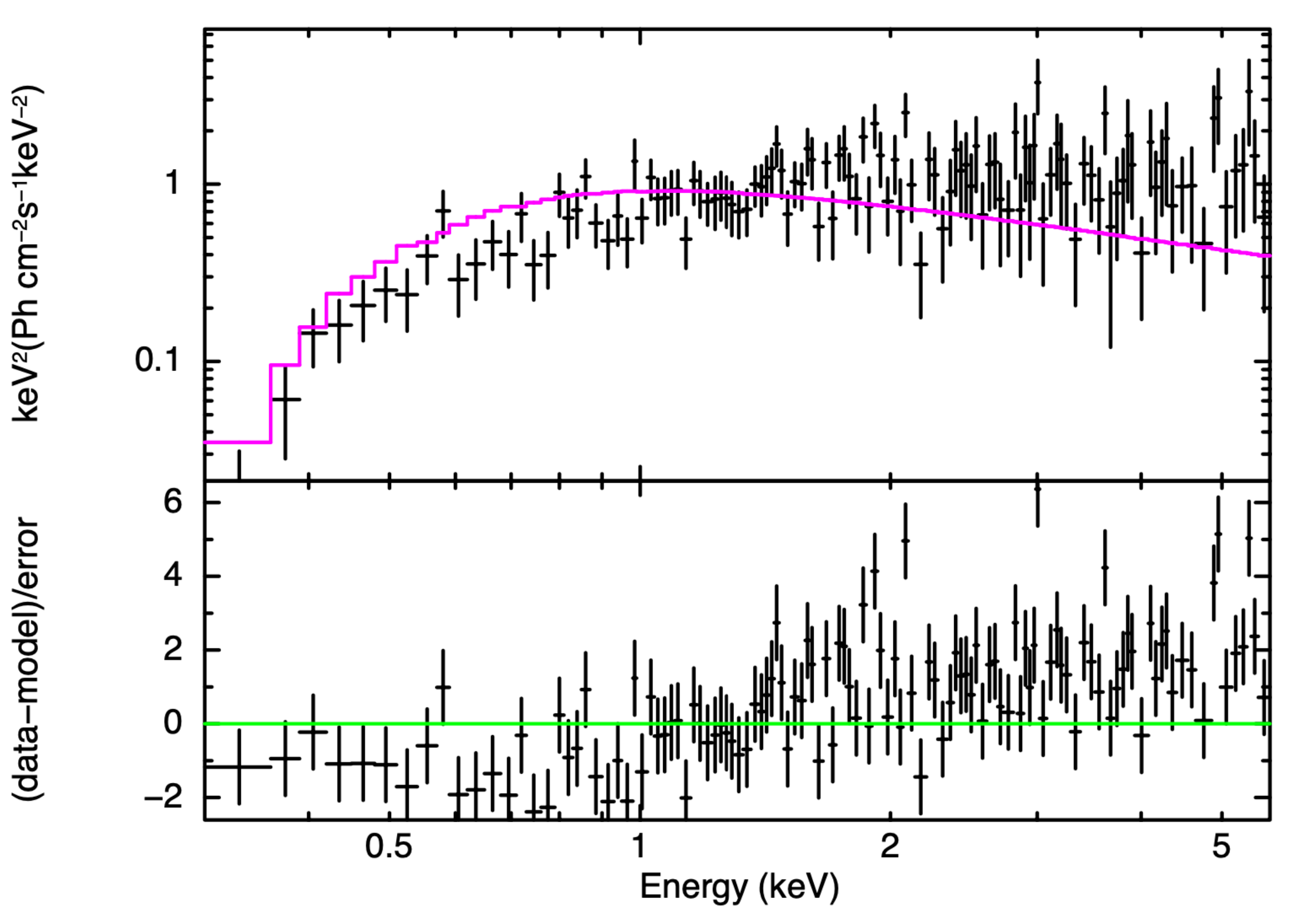}{0.5\textwidth}{}} 
\caption{Corner plots showing the posteriors for $N_{\rm{H,intr}}$ and $\Gamma$ for GRB~160117B (left) and the corresponding spectrum with the best-fit model (right). This GRB exhibits a decrease of $N_{\mathrm{H,intr}}$ with time, but was excluded from the final sample due to the strong spectral degeneracies and poor fits illustrated here. \label{fig:strongdegeneracies}}
\end{figure*}

For the selected GRBs, we finally investigated whether the observed decrease in $N_{\mathrm{H,intr}}$ may be driven by degeneracies between $N_{\mathrm{H,intr}}$ and $\Gamma$. This was done by inspecting the posterior distribution between $N_{\mathrm{H,intr}}$ and $\Gamma$. In Figure~\ref{fig:degeneracies}, we showcase two typical corner plots of posteriors: GRB~070306, where there is no degeneracy, and GRB~170607A, where $N_{\mathrm{H,intr}}$ and $\Gamma$ are degenerate. However, in the latter case the observed  degeneracies are not believed to cause the decay of $N_{\mathrm{H,intr}}$, since $\Gamma$ gets softer with time, opposite to the direction of the degeneracy. In this step, we also inspected the residuals of the best-fit models (shown in Figure \ref{fig:degeneracies}) to ensure that the data were well described by the models. 
After these inspections we discarded two GRBs: GRB~160117B, which showed very strong degeneracies between $N_{\mathrm{H,intr}}$ and $\Gamma$ as well as poor fits (illustrated in Figure~\ref{fig:strongdegeneracies}), and GRB~120811C, which showed spectral evolution in the same direction as the degeneracies. Of the remaining GRBs, five show no spectral degeneracies, while two show degeneracies like those presented in Figure \ref{fig:degeneracies} (corner plots and fits for all selected GRBs are included as online material associated with Figure~\ref{fig:degeneracies}).

\figsetstart
\figsetnum{2}
\figsettitle{Corner plots showing the posteriors for $N_{\rm{H,intr}}$ and $\Gamma$ (left) and the corresponding spectra with the best-fit model (right).}
\figsetgrpstart
\figsetgrpnum{2.1}
\figsetgrptitle{GRB~070306 corner plots}
\figsetplot{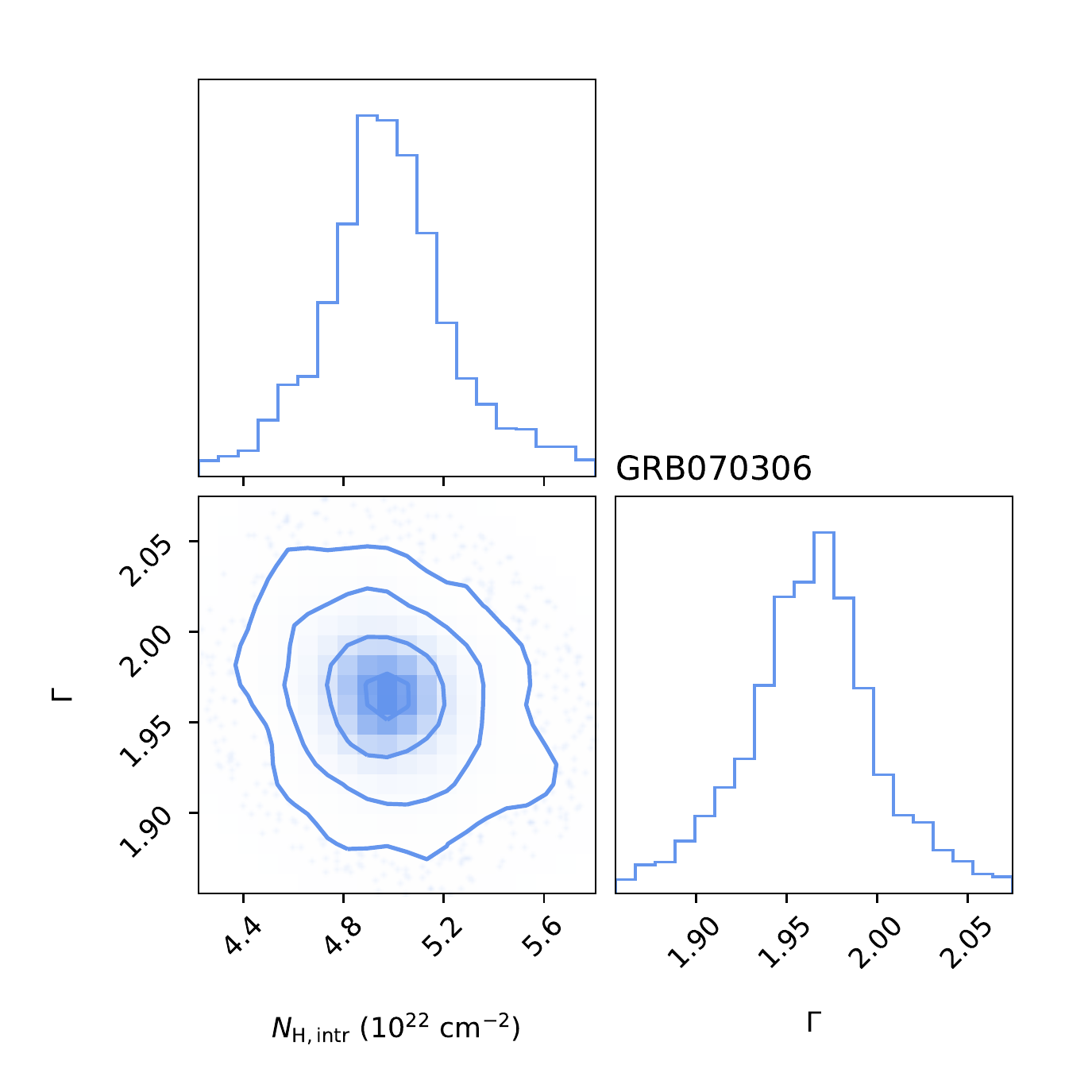}
\figsetgrpnote{Corner plots for GRB~070306.}
\figsetgrpend
\figsetgrpstart
\figsetgrpnum{2.2}
\figsetgrptitle{GRB~070306 best fits}
\figsetplot{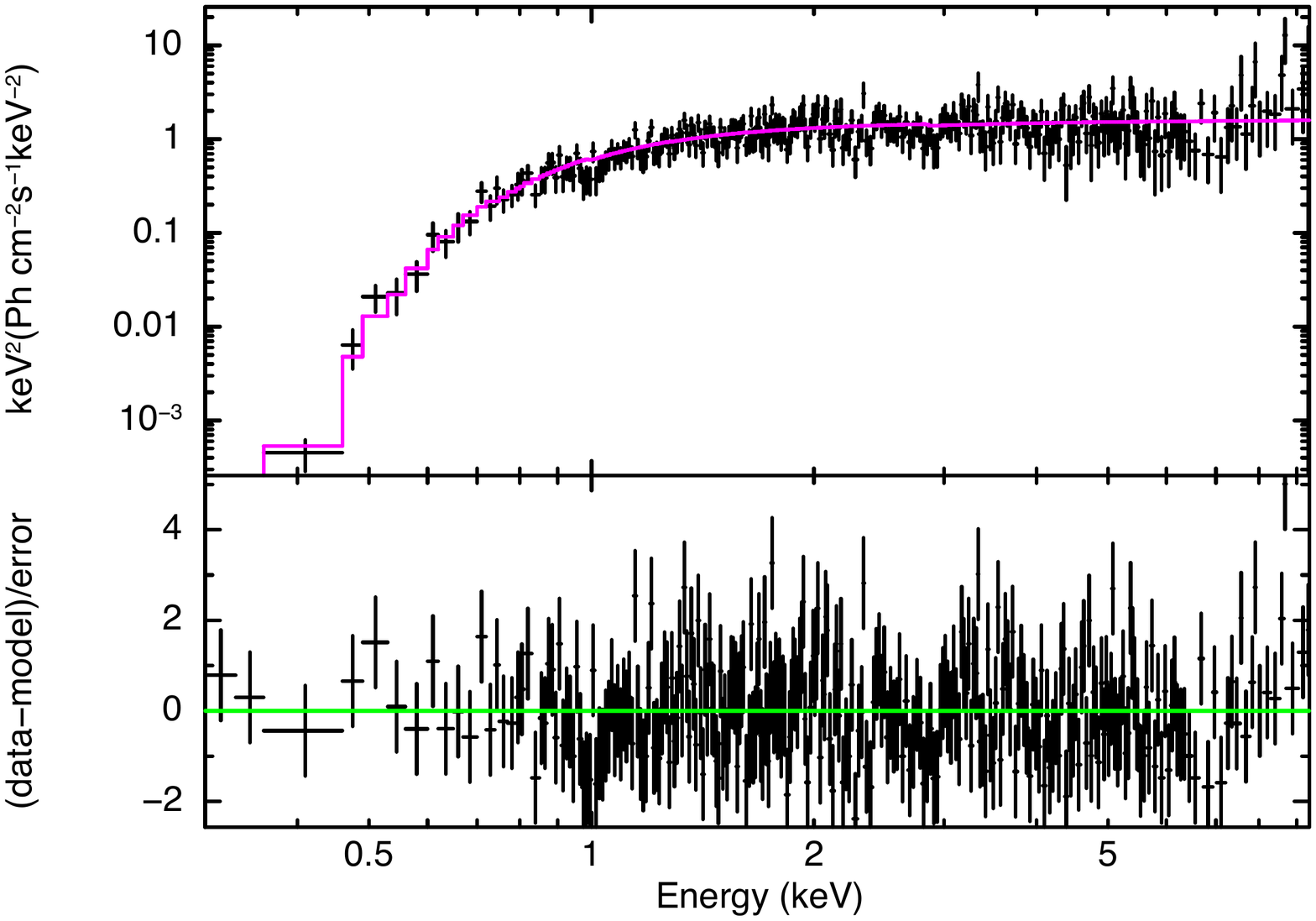}
\figsetgrpnote{Best fit plots for GRB~070306.}
\figsetgrpend
\figsetgrpstart
\figsetgrpnum{2.3}
\figsetgrptitle{GRB~071112C corner plots}
\figsetplot{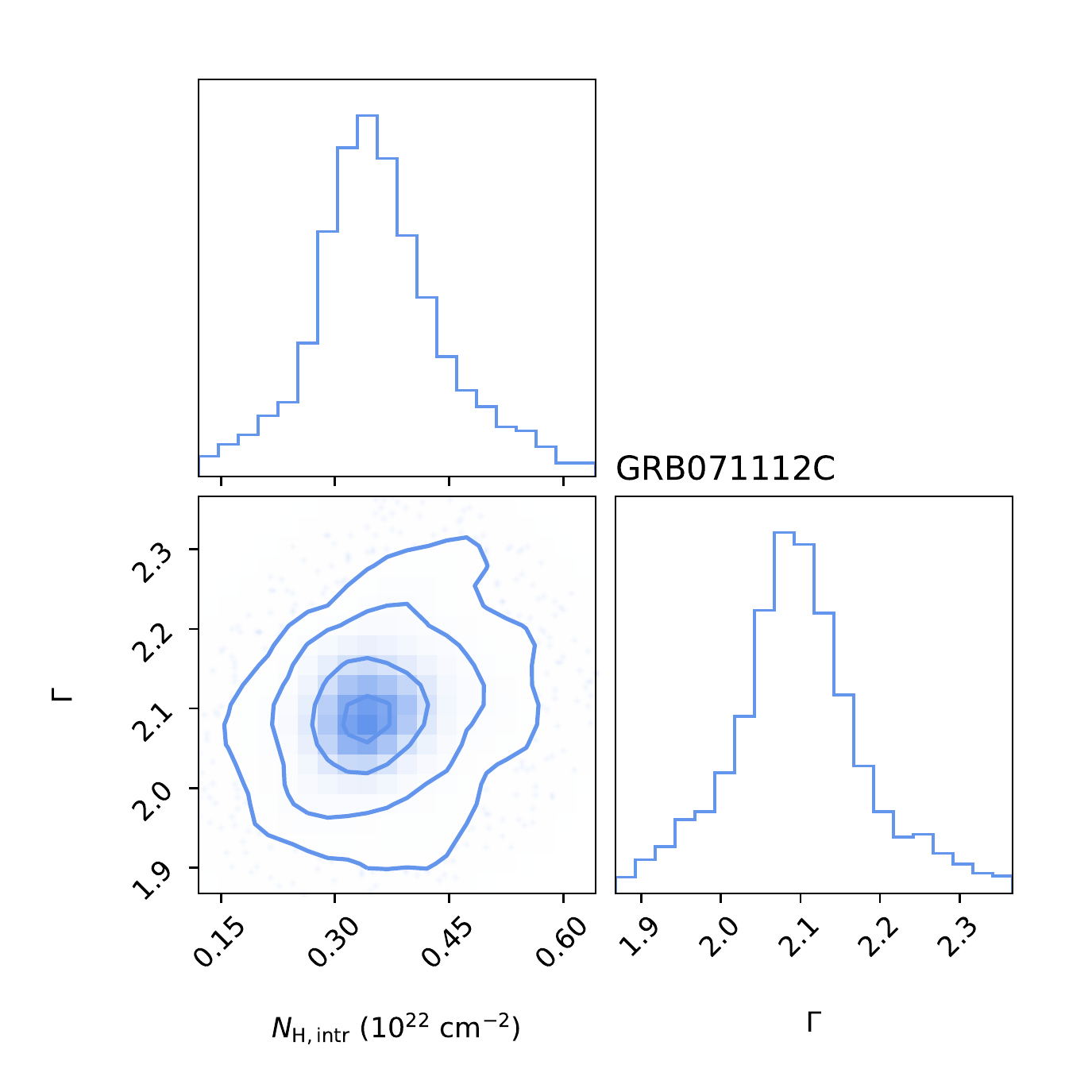}
\figsetgrpnote{Corner plots for GRB~071112C.}
\figsetgrpend
\figsetgrpstart
\figsetgrpnum{2.4}
\figsetgrptitle{GRB~071112C best fits}
\figsetplot{Online_material/GRB07111C_fits.pdf}
\figsetgrpnote{Best fit plots for GRB~071112C.}
\figsetgrpend
\figsetgrpstart
\figsetgrpnum{2.5}
\figsetgrptitle{GRB~081007 corner plots}
\figsetplot{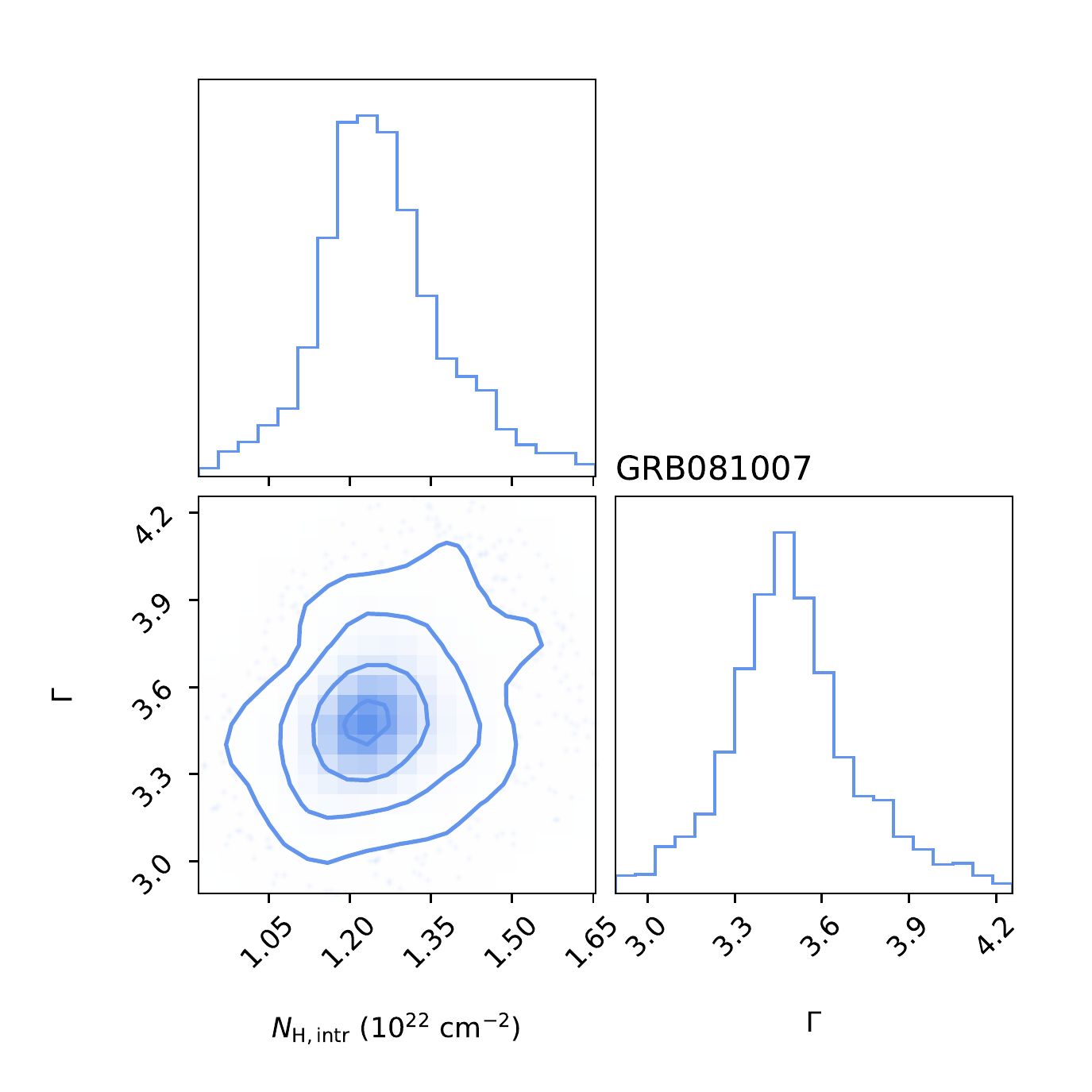}
\figsetgrpnote{Corner plots for GRB~081007.}
\figsetgrpend
\figsetgrpstart
\figsetgrpnum{2.6}
\figsetgrptitle{GRB~081007 best fits}
\figsetplot{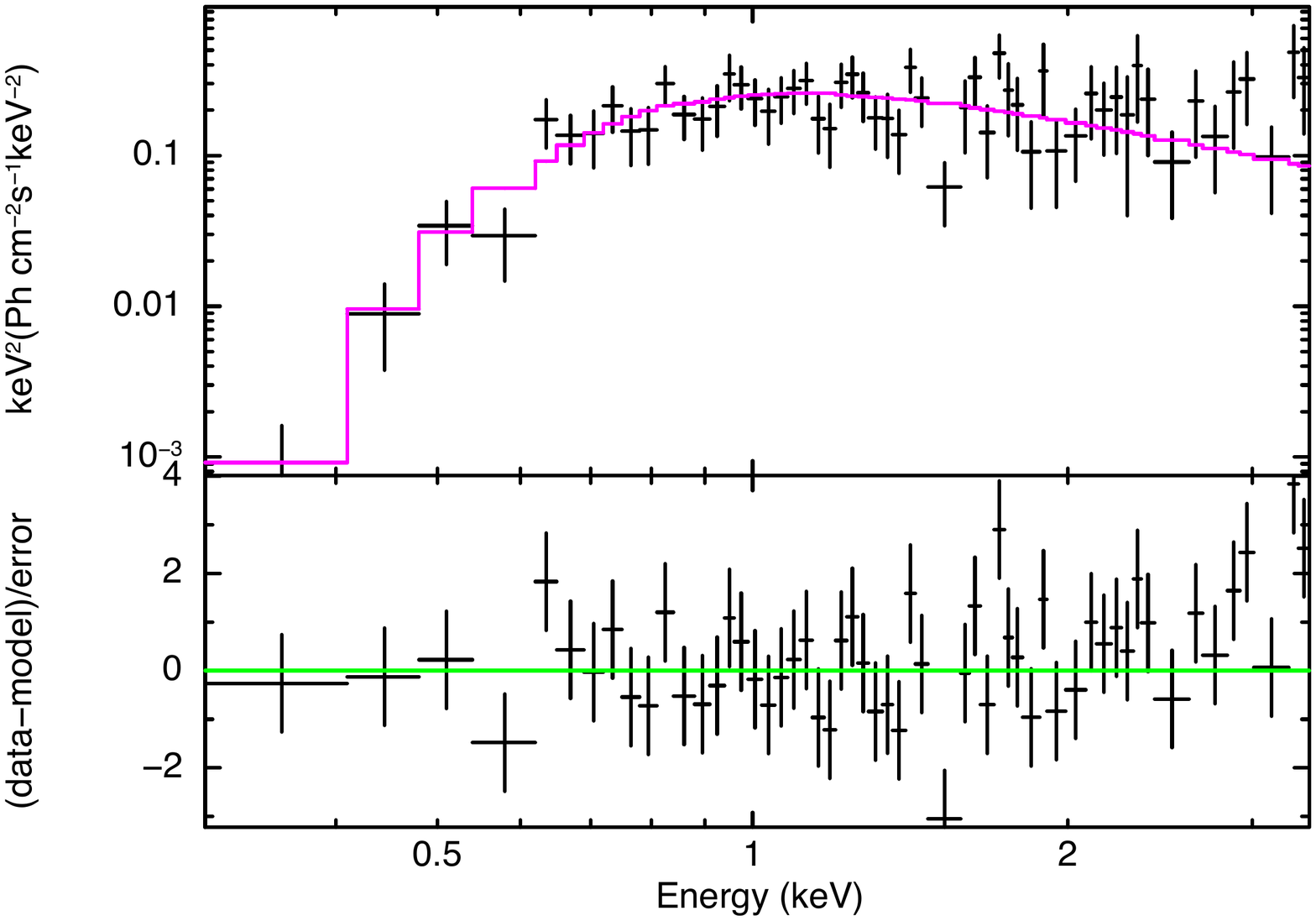}
\figsetgrpnote{Best fit plots for GRB~081007.}
\figsetgrpend
\figsetgrpstart
\figsetgrpnum{2.7}
\figsetgrptitle{GRB~090926B corner plots}
\figsetplot{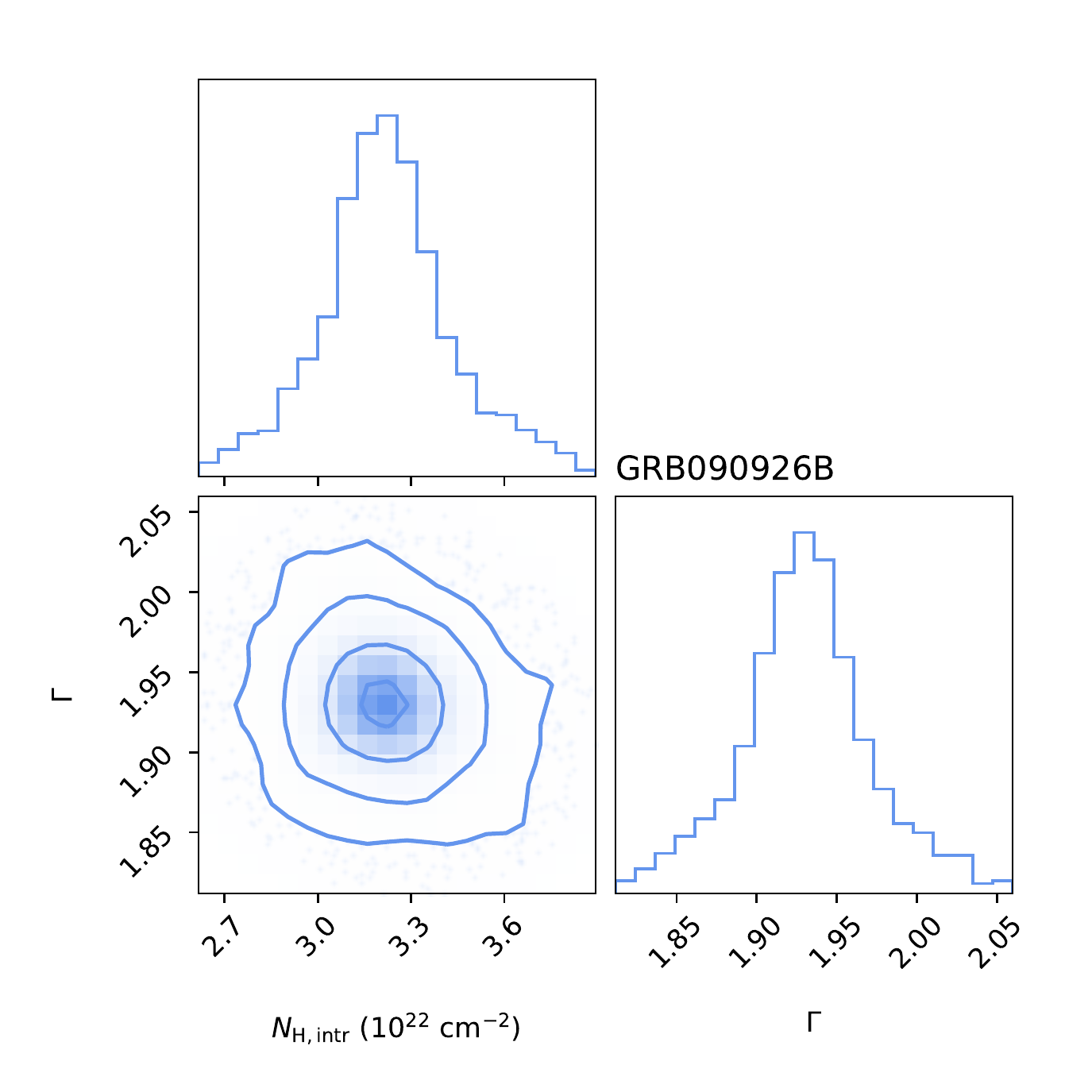}
\figsetgrpnote{Corner plots for GRB~090926B.}
\figsetgrpend
\figsetgrpnum{2.8}
\figsetgrptitle{GRB~090926B best fits}
\figsetplot{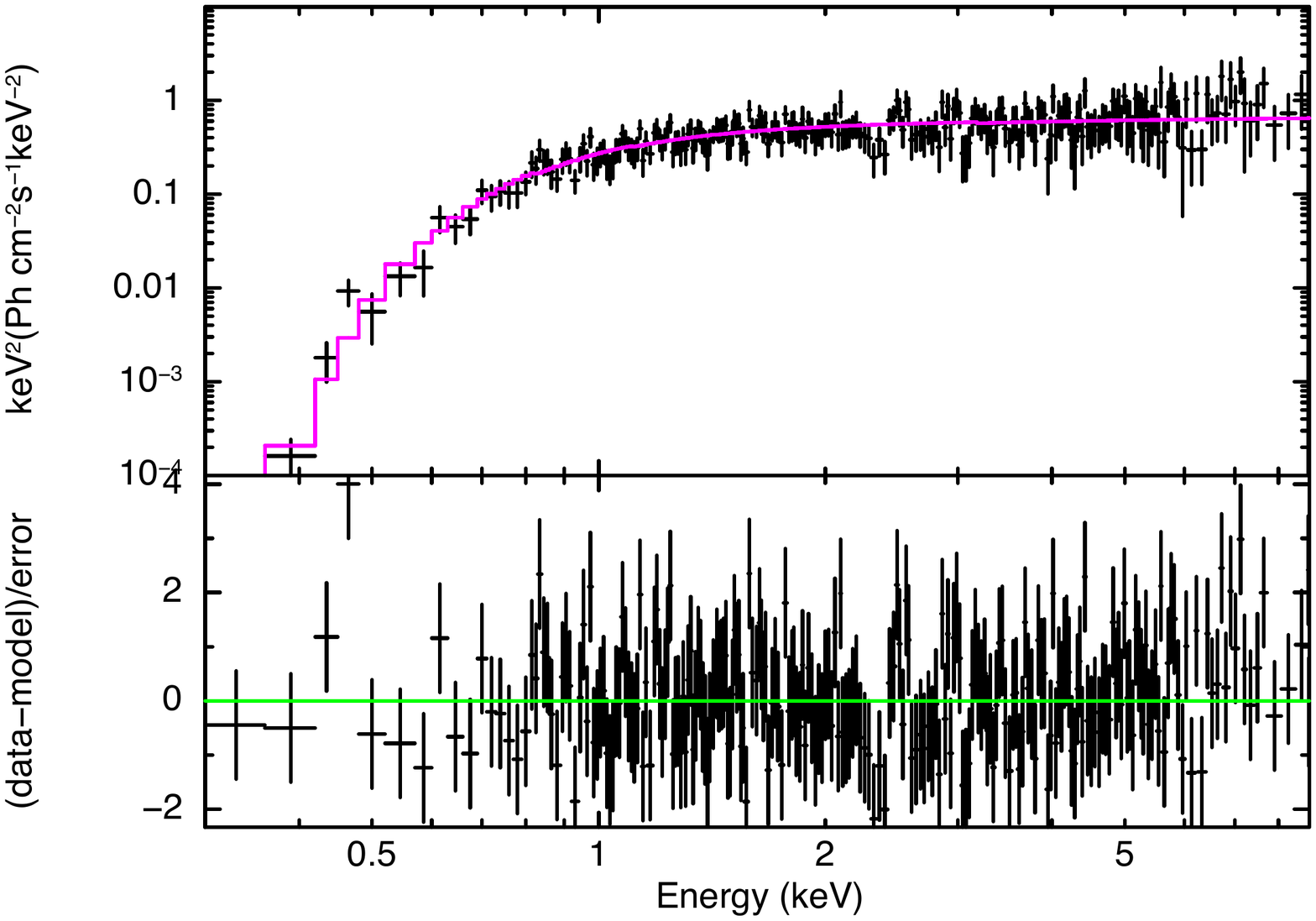}
\figsetgrpnote{Best fit plots for GRB~090926B.}
\figsetgrpend
\figsetgrpnum{2.9}
\figsetgrptitle{GRB~111125A corner plots}
\figsetplot{Online_material/GRB111125A_cornerplot.pdf}
\figsetgrpnote{Corner plots for GRB~111125A.}
\figsetgrpend
\figsetgrpnum{2.10}
\figsetgrptitle{GRB~111125A best fits}
\figsetplot{Online_material/GRB111126A_fits.pdf}
\figsetgrpnote{Best fit plots for GRB~111125A.}
\figsetgrpend
\figsetgrpnum{2.11}
\figsetgrptitle{GRB~170607A corner plots}
\figsetplot{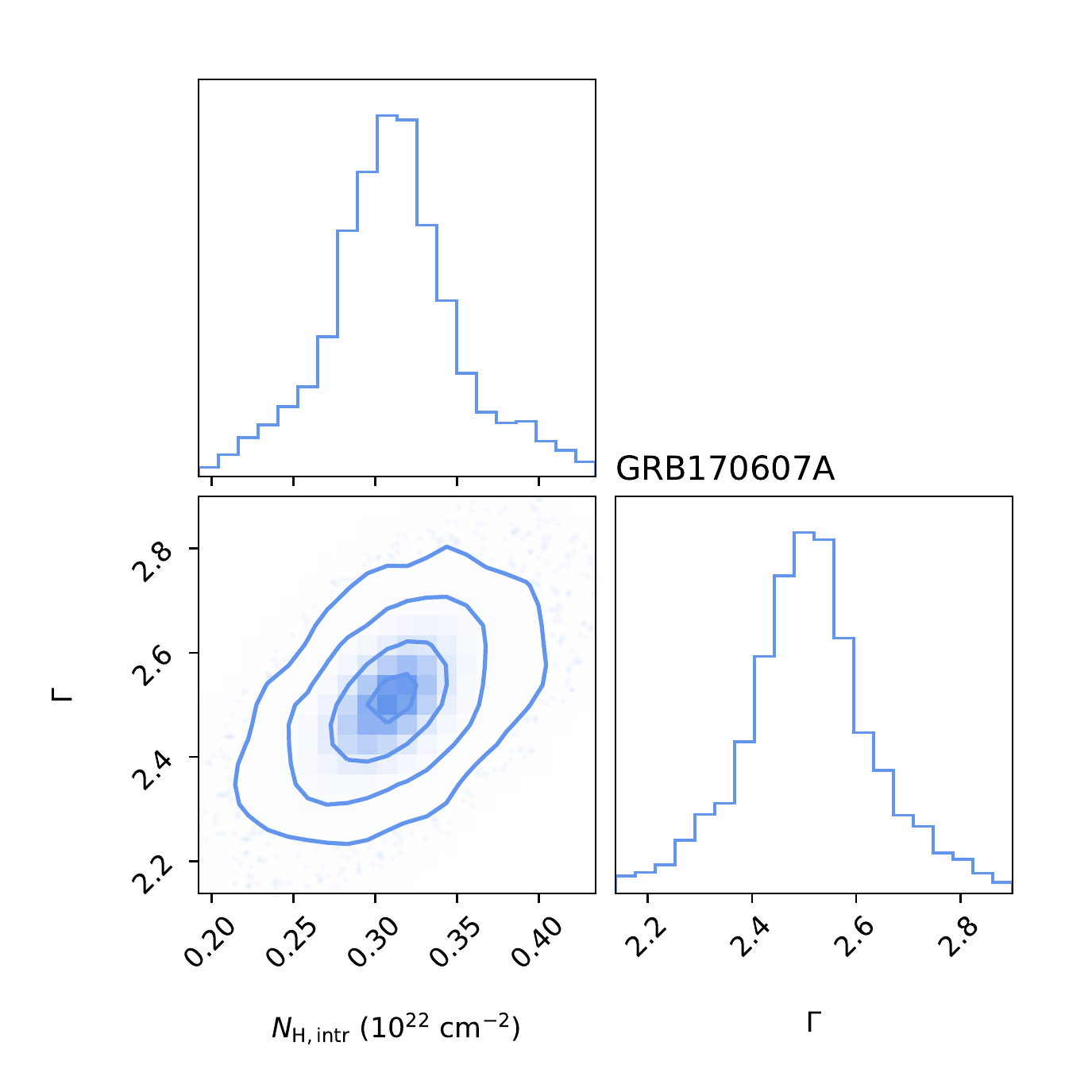}
\figsetgrpnote{Corner plots for GRB~170607A.}
\figsetgrpend
\figsetgrpnum{2.12}
\figsetgrptitle{GRB~170607A best fits}
\figsetplot{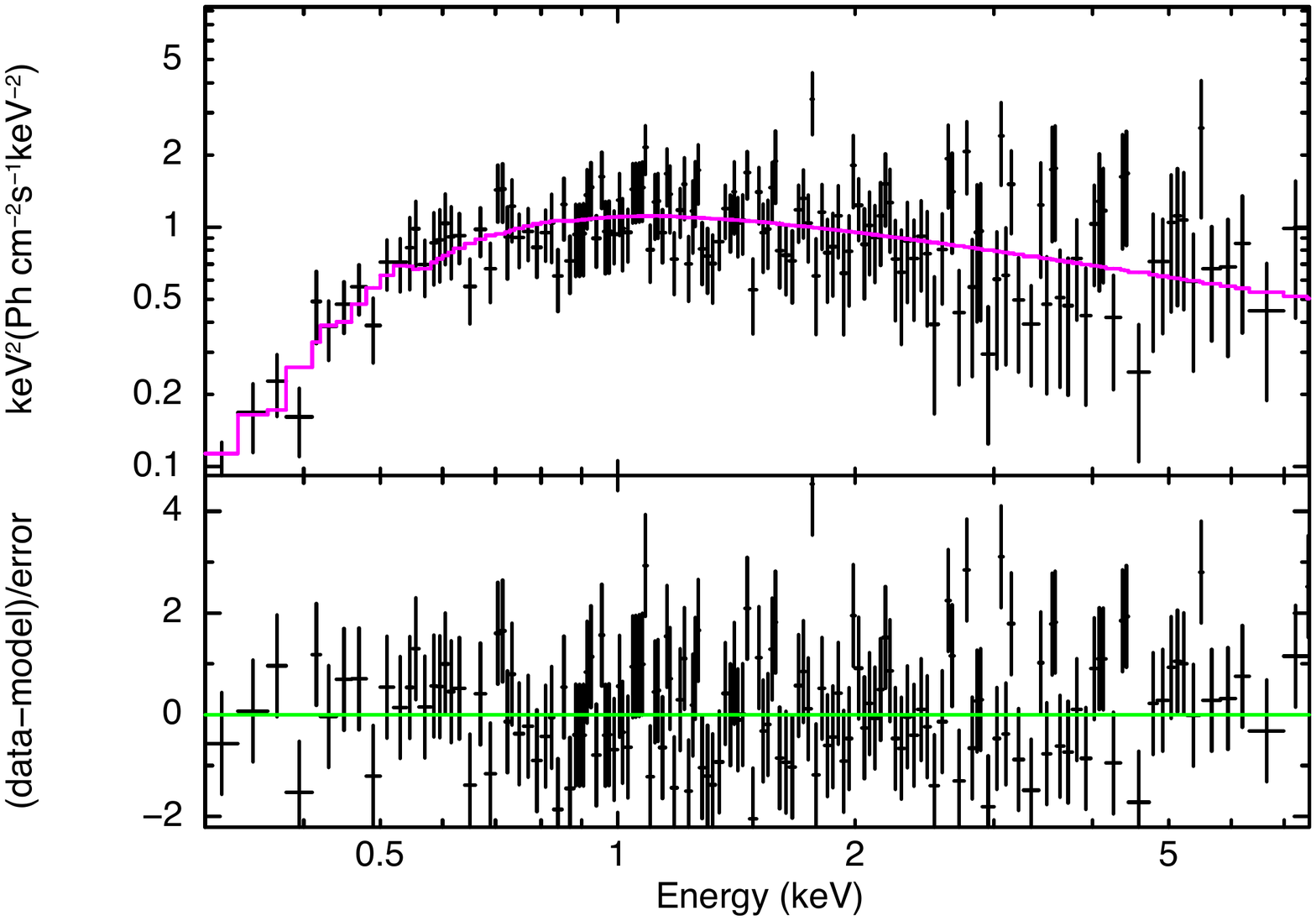}
\figsetgrpnote{Best fit plots for GRB~170607A.}
\figsetgrpend
\figsetgrpnum{2.13}
\figsetgrptitle{GRB~171222A corner plots}
\figsetplot{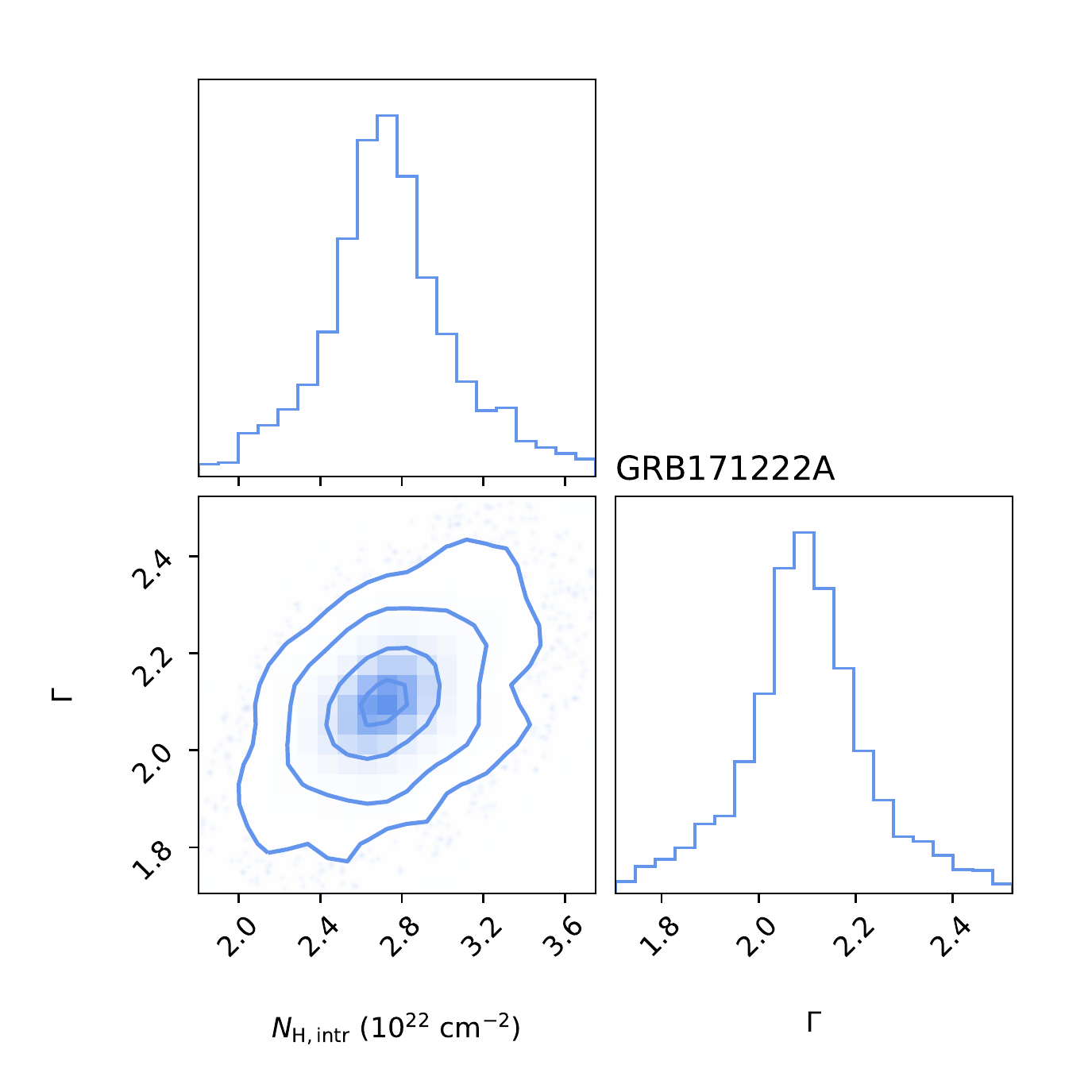}
\figsetgrpnote{Corner plots for GRB~171222A.}
\figsetgrpend
\figsetgrpnum{2.14}
\figsetgrptitle{GRB~171222A best fits}
\figsetplot{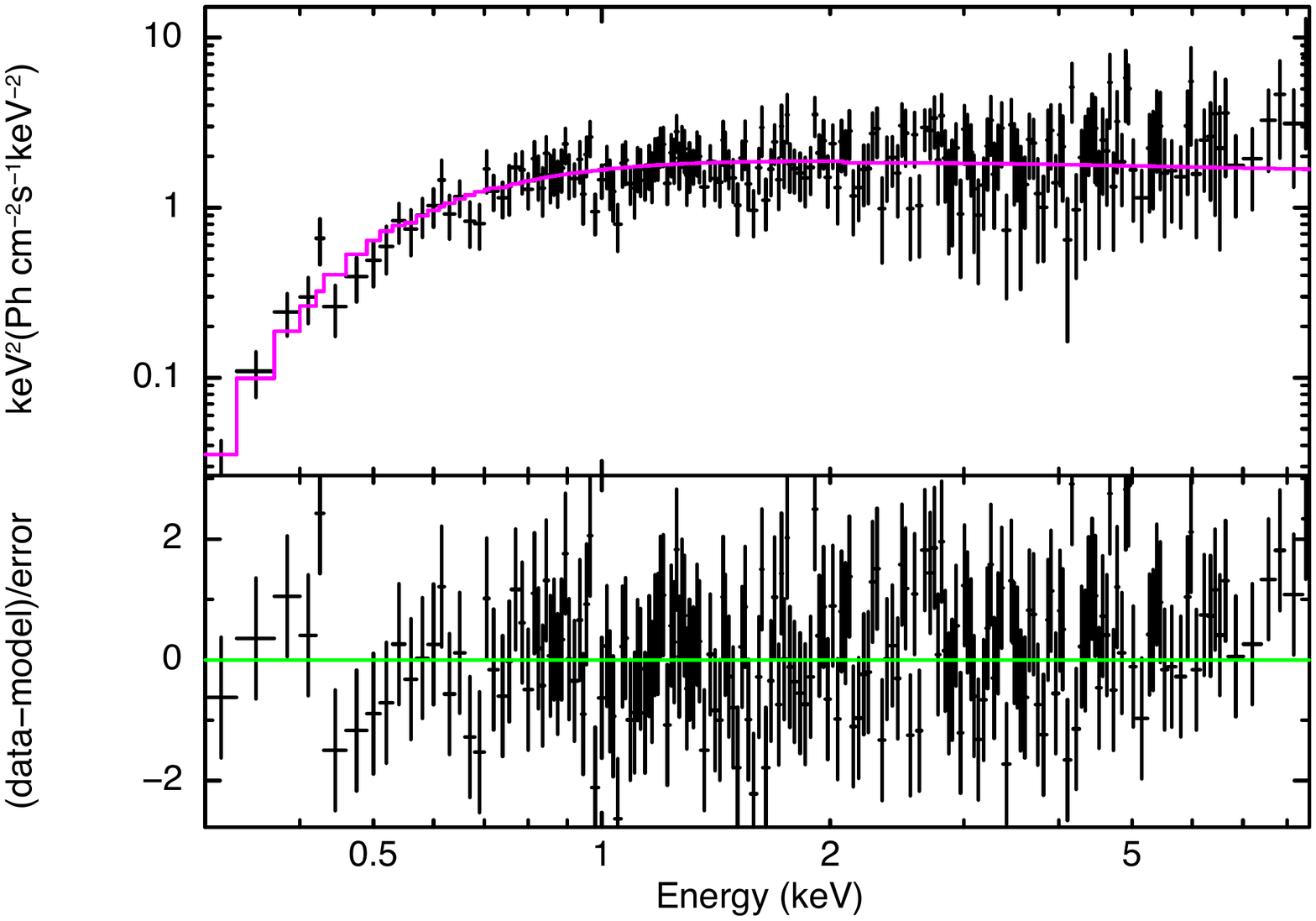}
\figsetgrpnote{Best fit plots for GRB~171222A.}
\figsetgrpend
\figsetend

\begin{figure*}[htb!]
\centering
    \includegraphics[width=\textwidth]{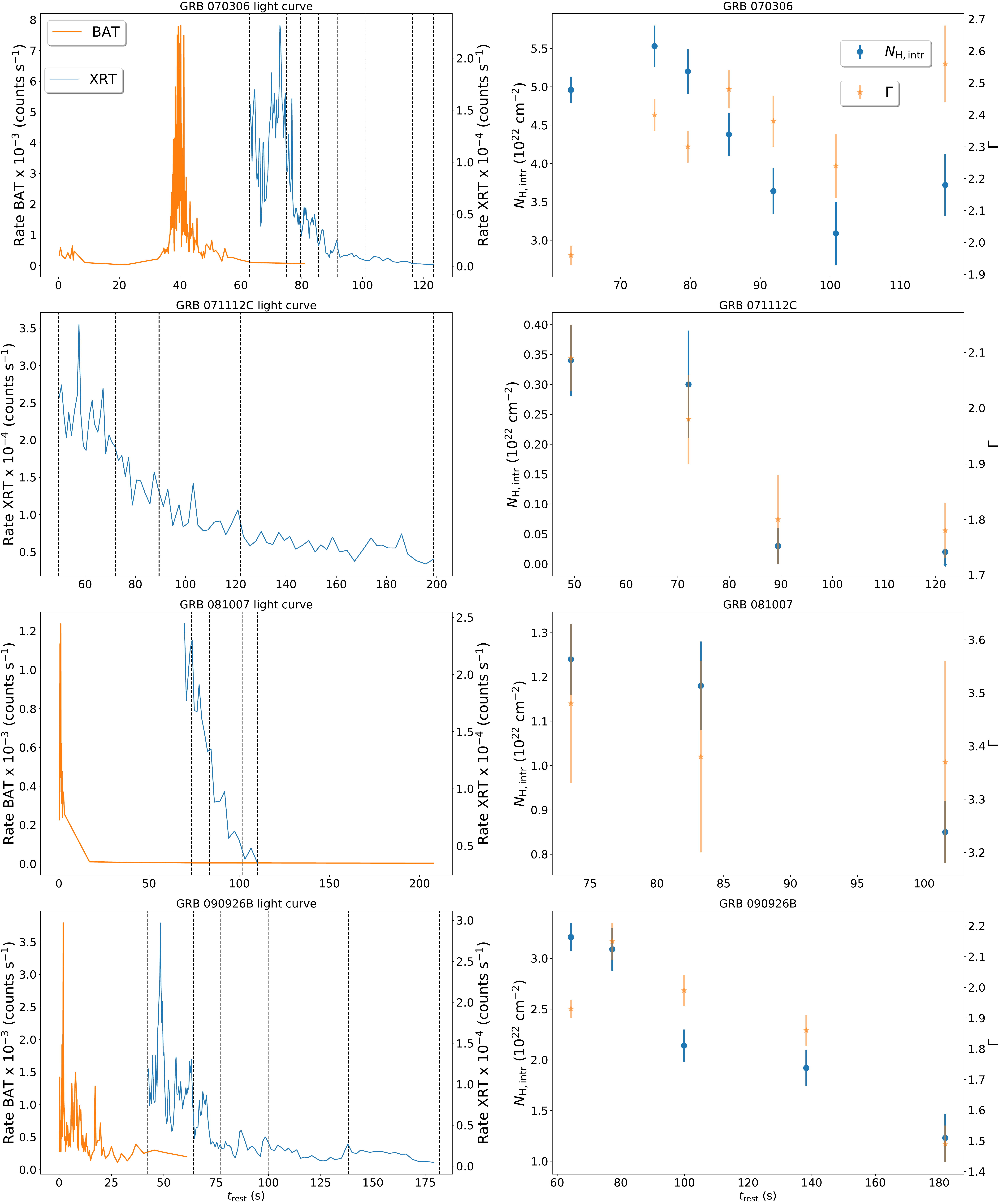}
\end{figure*}

\begin{figure*}[htb!]
\centering
    \includegraphics[width=\textwidth]{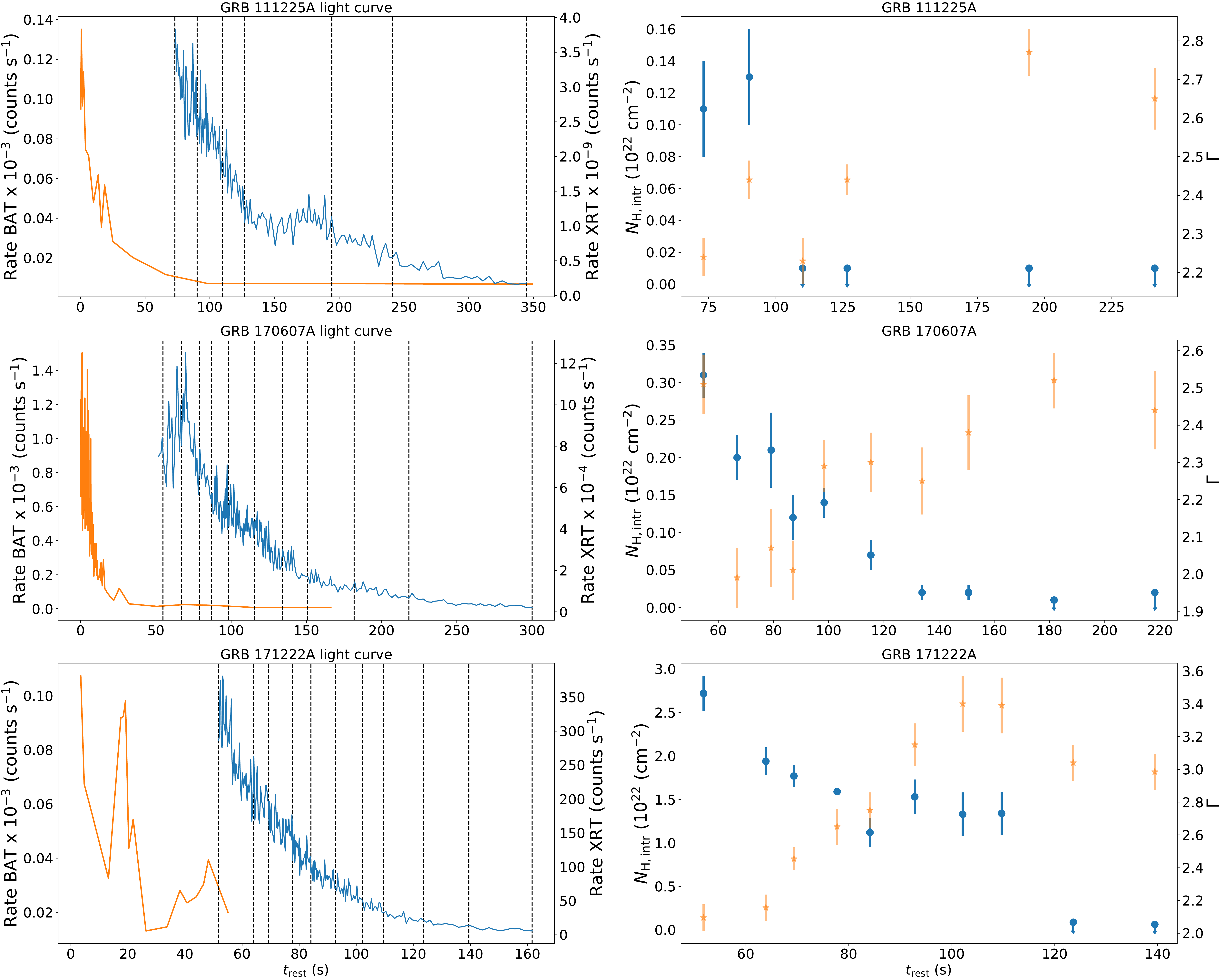}
\caption{Light curves (left) and time evolution of $N_{\mathrm{H,intr}}$ and $\Gamma$ (right) for the GRBs that show a decrease of $N_{\mathrm{H,intr}}$ (see also Table~\ref{tab:finalsample}). No BAT data was available for GRB~071112C. The dashed lines superposed on the light curves show the time intervals selected for the spectral analysis. $N_{\mathrm{H,intr}}$ is represented by blue circles and $\Gamma$ by orange stars. Note the different ranges on the y-axis for different GRBs.  \label{fig:Nh_timeevolution}}
\end{figure*}

\begin{deluxetable*}{c c c c c c c c c}[htb!]
\tablecaption{GRBs with possible decay of $N_\mathrm{H,intr}$. \label{tab:finalsample}}
\tablewidth{0pt}
\tablehead{
\colhead{GRB} & $z$  & $L_{\mathrm{av, 0.3-10keV}}$ & \colhead{$N_\mathrm{H,Gal}$} & $t_{N_\mathrm{H,intr,max}}$ -- $t_{N_\mathrm{H,intr,min}}$\tablenotemark{a}  & \colhead{$N_\mathrm{H,intr,max}$}  & \colhead{$N_\mathrm{H,intr,min}$} & \colhead{$N_\mathrm{H,intr,fixed}$\tablenotemark{b}} & \colhead{$N_\mathrm{H, PC}$\tablenotemark{c}} \\
\colhead{} & \colhead{} & \colhead{erg $\mathrm{s^{-1}}$} & \colhead{$10^{22} \ \mathrm{cm^{-2}}$}  & \colhead{s} & \colhead{$10^{22} \ \mathrm{cm^{-2}}$} & \colhead{$10^{22} \ \mathrm{cm^{-2}}$} & \colhead{$10^{22} \ \mathrm{cm^{-2}}$} & \colhead{$10^{22} \ \mathrm{cm^{-2}}$} } 
\startdata
070306 & 1.4959 & $2.36 \times 10^{49}$ & 0.0313 & 62.9 - 116.5 & $4.96 \pm 0.17$ & $3.72 \pm 0.40$ & $4.47 \pm 0.58$  & $3.60 \pm 0.45$   \\
071112C & 0.823 & $1.30 \times 10^{48}$  & 0.1110 &  49.3 - 121.8 & $0.34 \pm 0.06$ & $< 0.02$ & $ 0.01 \pm 0.001 $ & $ < 0.12$  \\
081007A & 0.5295 & $3.32 \times 10^{47}$ & 0.0144 & 73.6 - 101.6 & $1.24 \pm 0.08 $ & $0.85 \pm 0.07$ & $ 0.79 \pm 0.10 $  & $0.71 \pm 0.14$   \\
090926B & 1.24 & $6.19 \times 10^{48}$ & 0.0202 & 64.3 - 182.0  & $3.21 \pm 0.14$  & $1.23 \pm 0.24$ & $ 2.37 \pm 0.31 $  & $2.3 \pm 0.7$ \\
111225A* & 0.297 & $1.60 \times 10^{47}$ & 0.0172 & 70.1 - 241.0 & $0.11 \pm 0.03$ & $< 0.01 $ & $0.17 \pm 0.05$ & $< 0.15$ \\
170607A & 0.557 & $1.70 \times 10^{48}$ & 0.0437 &  54.7 - 218.1  & $0.31 \pm 0.03$ & $< 0.02$ & $ 0.05 \pm 0.001 $ & $0.04 \pm 0.02$ \\
171222A* & 2.409 & $1.15 \times 10^{50}$ & 0.0107 & 51.8 - 139.4 & $2.72 \pm 0.20$ & $< 0.06$ & $1.23 \pm 0.16$ & $<0.28$ \\
\enddata
\tablenotetext{a} {Time interval between the maximum and minimum values of $N_{\mathrm{H,intr}}$ obtained from the fits.} 
\tablenotetext{b} {$N_{\mathrm{H,intr,fixed}}$ is the value derived using the method in V18 (see also Section~\ref{sec:analysis}). Reported uncertainties correspond to the $90\%$ confidence interval.} 
\tablenotetext{c}{$N_{\mathrm{H,PC}}$ is the value obtained from the \textit{Swift} XRT catalogue.\footnote{\url{https://www.swift.ac.uk/xrt_live_cat/}} It is derived by fitting PC data and the reported uncertainties correspond to the $90\%$ confidence interval.}
* These GRBs have reported blackbody components in the XRT spectra (see V18, V21).
\end{deluxetable*}

The main properties of the seven selected GRBs are summarized in Table~\ref{tab:finalsample}, while Figure~\ref{fig:Nh_timeevolution} shows the light curves and time evolution of spectral parameters. It is clear that the significance of the decrease in $N_{\mathrm{H,intr}}$ is relatively low in some cases. The difference between the maximal and minimal $N_{\mathrm{H,intr}}$ reaches the 3$\sigma$-level in GRB~070306, GRB~090926B, GRB~170607A and GRB~171222A. However, we note that GRB~070306 also shows an increase in $N_{\mathrm{H,intr}}$ between the first and second time bins that is significant at $1.1 \sigma$, which somewhat weakens the case for this GRB. We also compared our results with the $N_{\rm{H,intr}}$ obtained from the data taken in photon counting (PC) mode after the WT observations analysed here had ended. These values ($N_{\rm H,PC}$) were obtained from the automated fits in the XRT live catalogue and are reported in Table~\ref{tab:finalsample}. We see that they are consistent with the minimal $N_{\rm{H,intr,min}}$ from our analysis (as well as with $N_{\rm{H,intr,fixed}}$) in all seven GRBs, showing that there is no evidence for a further decrease at late times considering the parameter uncertainties.

Two out of these seven GRBs (GRB~111225A and GRB~171222A) have previously been shown to have a blackbody component in their early X-ray spectra, under the assumption that the absorption is constant (V21). We begin by investigating the five GRBs without blackbody components in more detail. As described in Section~\ref{sec:analysis}, we fitted their spectra with both a power law and a cutoff power law with fixed $N_{\rm{H,intr}}$. The  Bayes factor between the power-law models with free and fixed $N_{\rm{H,intr}}$ are reported in Table~\ref{tab:bayesfactor}.
All GRBs have at least one time bin where the model with free $N_{\rm{H,intr}}$ is preferred according to the Bayes factor, which provides evidence that $N_{\rm{H,intr}}$ varies during the GRB. However, the model with fixed $N_{\rm{H,intr}}$ is strongly preferred in the majority of spectra, which illustrates that the variations are not significant between all time bins. This is also clear from the size of the confidence intervals on $N_{\rm{H,intr}}$ shown in Figure~\ref{fig:Nh_timeevolution}. In addition, given that $N_{\rm{H,intr,fixed}}$ was determined by fitting all spectra simultaneously (see Section \ref{sec:analysis}), it is expected that the model with $N_{\rm{H,intr,fixed}}$ will be preferred in some time bins even if the parameter constraints are tight. The strongest preference for the model with free $N_{\rm{H,intr}}$ is seen in the time bin with the highest $N_{\rm{H,intr}}$ in all GRBs.

Table~\ref{tab:bayesfactor} also lists the $\Delta$AIC between the power-law with free $N_{\rm{H,intr}}$ and the cutoff power law with $N_{\rm{H,intr,fixed}}$. 
The only GRB that shows strong preference for the power-law model in all bins is GRB~090926B. In all other cases the results are inconclusive, with the preference for the two models changing with time, though GRB~170607A shows strong preference for the power-law in most bins, while GRB~070306 shows strong preference for the cutoff power-law in most bins.
The best-fit parameters of the cutoff power law are provided in Appendix \ref{appendix:bayesfactor}. It is notable that the values of $E_{\rm cut}$ evolve erratically, with many values being above 20~keV, meaning that the curvature has a very small impact in the fitted energy range.

For the two GRBs that have reported blackbody components in their spectra, we have calculated the $\Delta$AIC between the power-law plus blackbody and power-law with free $N_{\rm{H,intr}}$ models. The results are presented in Table 3. For GRB 111225A, they show strong support for the power-law plus blackbody model in all but the first two bins. For GRB~171222A, the magnitude of  $\Delta$AIC is smaller and it alternates between the two models. However, we note that when it prefers the power-law plus blackbody model, the preference is stronger. Based on these results and inspection of the fits (poor power-law fits in, e.g., bin 3 in GRB~111225A and bin 5 in GRB~171222A, see online material associated with Figure~\ref{fig:degeneracies}), we give preference to the power-law plus blackbody model for these two GRBs, though the case is clearly weaker for GRB~171222A.

\begin{deluxetable*}{cccccc}[htb!]
\tablecaption{Comparison of different models fitted to the GRBs without blackbody components in Table~\ref{tab:finalsample}.
\label{tab:bayesfactor}}
\tablewidth{0pt}
\tablehead{
\colhead{GRB} & \colhead{time bin\tablenotemark{a}} & \colhead{Log evidence PL\tablenotemark{b}}  & \colhead{Log evidence PLf\tablenotemark{c}} & \colhead{$\mathrm{log (BF_{PL-PLf})}$\tablenotemark{d}} & \colhead{$\mathrm{AIC_{CPL} - AIC_{PL}}$\tablenotemark{e}}\\
\colhead{} &\colhead{} & \colhead{power-law, free $N_\mathrm{H,intr}$} & \colhead{power-law, fixed $N_\mathrm{H,intr}$} & \colhead{} & \colhead{}} 
\startdata
070306 & bin1 & -255.12 &  -252.22 & -2.90 & 498.79\\
070306 & bin2 & -165.12 &  -167.39 & 2.27 & -16.17\\
070306 & bin3 & -168.90 &  -167.95 &  -0.95 & 39.37\\
070306 & bin4 & -158.62 &  -149.73 & -8.89 & -15.06\\
070306 & bin5 & -147.39 &  -141.01 & -6.38 & -35.57\\
070306 & bin6 & -91.36 &  -87.02 & -4.34 & -31.03\\
070306 & bin7 & -87.00 &  -84.07 & -2.93 & -25.92\\
\hline
071112C & bin1 & -166.49 & -170.54 & 4.05 & 23.77\\
071112C & bin2 & -142.43 & -142.24 & -0.19 & -30.73\\
071112C & bin3 & -129.10 & -124.00 & -5.10 & 1.69\\
071112C & bin4 & -196.45 & -191.95 & -4.5 & -1.25\\
\hline
081007 & bin1 & -90.42 & -91.32 & 0.90 & -1.25\\
081007 & bin2 & -59.67 & -58.89 & -0.78 & 1.85\\
081007 & bin3 & -79.77 & -76.69 & -3.08 & -2.31\\
\hline
090926B & bin1 & -257.95 & -260.64 & 2.69 & 87.99\\
090926B & bin2 & -166.54 & -166.22 & -0.32 & 20.88\\
090926B & bin3 & -166.96 & -165.04 & -1.92 & 31.60\\
090926B & bin4 & -171.04 & -169.76 & -1.28 & 35.62\\
090926B & bin5 & -113.31 & -114.83 & 1.52 & 26.57\\
\hline
170607A & bin1 & -106.04 & -114.11 & 8.07 & 48.76\\
170607A & bin2 & -107.26 & -107.31 & 0.05 & 26.74\\
170607A & bin3 & -86.34  & -84.72 & -1.62 & -144.79\\
170607A & bin4 & -107.11 & -103.97 & -3.14 & -40.66\\
170607A & bin5 & -121.49 & -119.67 & -1.82 & 33.49\\
170607A & bin6 & -125.09 & -120.89 & -4.20 & 90.78\\
170607A & bin7 & -119.77 & -115.56 & -4.21 & 72.26\\
170607A & bin8 & -96.09  & -91.82 & -4.27 & 60.47\\
170607A & bin9 & -109.76 & -105.04 & -4.72 & 14.95\\
170607A & bin10 & -0.97 & -0.96 & -0.01 & 1.95\\
\enddata
\tablenotetext{a}{The time bins are plotted in Figure \ref{fig:Nh_timeevolution}.}
\tablenotetext{b}{Log evidence for the power-law model with free $N_\mathrm{H,intr}$.}
\tablenotetext{c}{Log evidence for the power-law model with fixed $N_\mathrm{H,intr}$.}
\tablenotetext{d}{Bayes factor (Equation \ref{eq:BF}) between the models in the previous columns. Positive values indicate that the power-law model with free $N_\mathrm{H,intr}$ is preferred. A GRB with a constant  $N_{\rm{H,intr}}$ is expected to have negative values in all time bins.}
\tablenotemark{e}{Difference between AIC for the cutoff power-law with fixed $N_\mathrm{H,intr}$ and the power-law with free $N_\mathrm{H,intr}$, as defined by Equation~\ref{eq:AIC1}. Positive values indicate that the power-law model with free $N_\mathrm{H,intr}$ is preferred.}
\end{deluxetable*}

\begin{deluxetable}{ccc}[htb!]
\tablecaption{Comparison of different models fitted to the GRBs with blackbody component in Table~\ref{tab:finalsample}.
\label{tab:AIC}}
\tablewidth{0pt}
\tablehead{
\colhead{GRB} & time bin\tablenotemark{a} & \colhead{$\mathrm{AIC_{BB} - AIC_{PL}}$\tablenotemark{b}}\\
\colhead{} &\colhead{} & \colhead{}} 
\startdata
111225A & bin 1 & 3.6 \\
111225A & bin 2 & -2.32 \\
111225A & bin 3 & -23.85 \\
111225A & bin 4 & -41.95 \\
111225A & bin 5 & -64.31 \\
111225A & bin 6 & -24.48 \\
\hline
171222A & bin 1 & 1.38 \\
171222A & bin 2 & 2.18 \\
171222A & bin 3 & -5.73 \\
171222A & bin 4 & 0.94 \\
171222A & bin 5 & -7.74 \\
171222A & bin 6 & -11.8 \\
171222A & bin 7 & -13.19 \\
171222A & bin 8 & -0.28 \\
171222A & bin 9 & 1.81 \\
171222A & bin 10 & -17.35 \\
\enddata
\tablenotetext{a}{Time bins corresponding to the ones plotted in Figure \ref{fig:Nh_timeevolution}.}
\tablenotemark{b}{Difference between AIC for power-law plus blackbody and power-law with free $N_\mathrm{H,intr}$ as defined by Equation \ref{eq:AIC1}. If the difference is negative, the power-law plus blackbody model is preferred.}
\end{deluxetable}

\section{Discussion} \label{sec:discussion}

Our results show that only seven GRBs out of 199 analyzed in total, corresponding to $3.5\%$ of the sample, exhibit some evidence of a decreasing $N_{\mathrm{H,intr}}$. Out of these seven, two GRBs have previously reported blackbody components in their early X-ray spectra. On the whole, the evidence for a decrease in absorption in these seven GRBs is tentative. While inspection of the posteriors show that the decrease is very unlikely to be caused by spectral degeneracies between $N_{\mathrm{H,intr}}$ and $\Gamma$, it is clear that the significance of the decrease is modest in many cases and that alternative models cannot be completely ruled out. 

The strongest evidence is seen in GRB~090926B, where the decrease of $N_{\mathrm{H,intr}}$ is at the $3\sigma$ level and the $\Delta \mathrm{AIC}$ clearly favours the power-law model with free $N_{\mathrm{H,intr}}$ over the cutoff power law with fixed $N_{\mathrm{H,intr}}$. On the other side is GRB~111225A, which represents the case where it is least probable that the observed decrease in $N_{\mathrm{H,intr}}$ is real. In the majority of bins, there is only an upper limit on $N_{\mathrm{H,intr}}$ and the $\Delta \mathrm{AIC}$ favours the power-law plus blackbody model with fixed $N_{\mathrm{H,intr}}$. 
The conclusions are less clear for the rest of the sample. 
In the other GRB with a possible blackbody component, GRB~171222A, $\Delta \mathrm{AIC}$ alternates between the power law with free $N_{\mathrm{H,intr}}$ and the power law plus blackbody with fixed $N_{\mathrm{H,intr}}$. This GRB also shows rather unusual blackbody properties with a high luminosity of $L_\mathrm{BB} \sim 10^{49} \ \mathrm{erg \ s^{-1}}$ together with a low temperature of $0.2 \le T_{\mathrm{BB}} \le 0.5$~ keV (see V21). Taking all this into account, we cannot exclude the possibility  that the inferred blackbody component is an artefact of  a decreasing $N_{\mathrm{H,intr}}$. For the remaining four GRBs, the alternative model of a cutoff power law with fixed $N_{\mathrm{H,intr}}$ provides a better fit in some time bins, but, on the other hand, does not show a systematic evolution of $E_{\rm cut}$.

If present, the decrease in $N_{\mathrm{H,intr}}$ represents a powerful tool in understanding the local environment of GRBs. Numerical simulations show that such a decrease is a natural consequence of ionization of the local medium by the GRB radiation \citep{Perna1998,Lazatti2002,Perna2002}. The ionization time scale is shorter if the luminosity of the GRB is high and the absorber is compact. This implies that any wind medium is expected to be ionised on a very short time scale of milliseconds \citep{Lazatti2002}, long before the start of the {\it Swift}~XRT observations. For the typical time scales probed in our analysis, we can instead expect to detect decreasing absorption from dense pc-scale clouds or shells, where the latter results in a slower decrease as all the material is located at a large distance \citep{Lazatti2002}. Shell-geometries may reflect H~II regions surrounding the massive progenitors.  

For the seven GRBs that show a possible decrease in $N_{\mathrm{H,intr}}$, the initial column densities are in the range $\sim 0.1- 5 \times \mathrm{10^{22} \ cm^{-2}}$ at the start of the observations around $t_{\rm rest} \sim 50$~s (Table~\ref{tab:finalsample}). Comparing to the simulations for a pc-size uniform cloud ionized by a constant luminosity presented in \citet{Lazatti2002}, the $N_{\mathrm{H,intr}}$ would have been $\gtrsim 5$ times higher at the time of the trigger. In this context it is interesting to note that GRB~070306 triggered on a weak precursor, but had its main emission episode in BAT at $\sim 20$~s before the start of the XRT observations (Figure~\ref{fig:Nh_timeevolution}). This GRB also has the highest initial $N_{\mathrm{H,intr}} \sim 5 \times \mathrm{ 10^{22} \ cm^{-2}}$,  which is compatible with the shorter time scale probed. 

The magnitude of the decrease in the seven GRBs over the observed 50--100~s spans a broad range, from a moderate $\sim 30\%$ to reaching non-detectable values of $N_{\mathrm{H,intr}} \lesssim \mathrm{ 10^{20} \ cm^{-2}}$. A detailed modelling of the individual GRBs, considering the light curves and spectra of the ionizing radiation, would need to be performed to assess if this is consistent with ionization of the local medium and to constrain the geometry and density of the possible absorbers. 

It is clear from our results that a measurable decreasing $N_{\mathrm{H,intr}}$ is very rare and that the only possible detections are at relatively low redshift (all GRBs in Table~\ref{tab:finalsample} have $z<2.5$). This is consistent with the scenario that the total $N_{\mathrm{H,intr}}$ is dominated by the IGM, but also with an origin in the GRB host galaxies themselves, as long as most of the absorbing material is not in the close vicinity of the GRBs. Models considering IGM absorption predict that is should start dominating over the host contribution at $z \sim 2-3$ \cite[e.g.,][]{Behar2011, Starling2013, Dalton2021}. This is consistent with the fact that we do not identify any GRBs with decreasing $N_{\mathrm{H,intr}}$ at high $z$, though it should be noted that the redshift of the spectrum also makes the detection of absorption at high $z$ more challenging.

The fact that $N_{\mathrm{H,intr}}$ is consistent with being constant at early times in the vast majority of GRBs also implies that estimates of $N_{\mathrm{H,intr}}$ obtained from WT data as in V18 and V21 (see also Section \ref{sec:analysis}) are suitable for investigating the correlation with redshift. Previous studies of the $N_{\mathrm{H,intr}} -z$ relation have instead used $N_{\mathrm{H,intr}}$ as determined from the lower count-rate late PC data to limit the effect of time evolution of the spectrum \cite[e.g.,][]{Starling2013, Rahin2019}. 

We have previously shown, based on a small sample in V18, that our values of $N_{\mathrm{H,intr}}$ from WT data are mostly consistent with those determined from late-time PC data, but typically better constrained due to the higher count rate during the WT mode (see Figure 1~in V18). Here we extend the comparison to the full sample using the XRT catalogue, which is based on 
\citet{Evans2009}. We have taken the values of $N_{\mathrm{H,intr}}$ derived from PC data for the same 199 GRBs used in this analysis. The fits in the catalogue were performed with the host redshift, and the reported uncertainties on $N_{\mathrm{H,intr}}$ correspond to the $90\%$ confidence interval. We find that the relative uncertainty on $N_{\mathrm{H,intr}}$ derived from PC data is on average $\sim 60\%$, with a non-negligible number of bursts having only upper limits on $N_{\mathrm{H,intr}}$. On the other hand, the relative 90\% uncertainties on $N_{\mathrm{H,intr}}$ derived from WT data using Bayesian blocks binning is on average $\sim 23 \%$. 

\begin{figure}[htb!]
    \centering
    \includegraphics[width=\columnwidth]{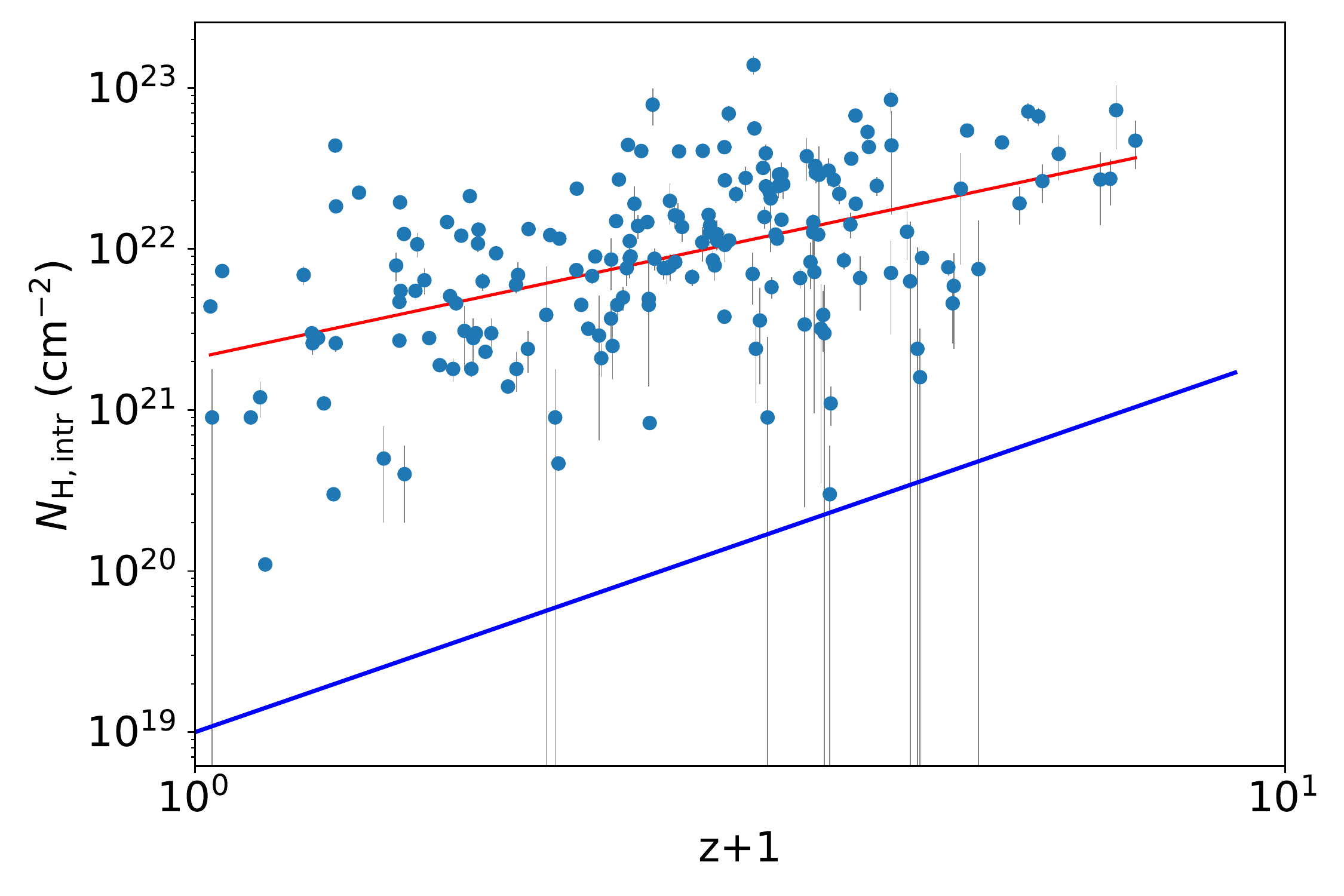}
    \caption{Relation between $N_{\mathrm{H,intr}}$ and redshift for the sample of 180 GRBs observed by \textit{Swift}. The red line represents the fit to the data and the blue line illustrates how the minimum detectable column density scales with redshift, $N_{\mathrm{H,intr}}$ = $N_{\mathrm{H,z=0}} \times (z+1)^{2.34}$ from \citet{Campana2014} for $N_{\mathrm{H,z=0}} = 10^{19} \mathrm{cm^{-2}}$.}
    \label{fig:Nh-z}
\end{figure}

We also investigate a possible correlation between $N_{\mathrm{H,intr}}$ and $z$ with our data set and find it present as shown in Figure~\ref{fig:Nh-z} (where the 19 GRBs with blackbody components identified in V21 have been excluded). Fitting the data using \textsc{scipy.stats.linregression} gives  $N_{\mathrm{H,intr}} \propto (z+1)^{1.44 \pm 0.19}$, which is consistent with previous results based on PC data. \citet{ Rahin2019} reported $N_{\mathrm{H,intr}} \propto (z+1)^{1.5}$ over the entire redshift range. \citet{Dalton2020} also investigated the observed correlation, but have not reported any scaling coefficient between $N_{\mathrm{H,intr}}$ and $z$. However, a comparison of our results with Figure 1 in \citet{Dalton2020} shows $N_{\mathrm{H,intr}}$ spanning the same range of values.

The fact that a decrease of $N_{\mathrm{H,intr}}$ is very rare also means that this does not interfere with modelling of the intrinsic spectrum in the late prompt emission and/or early afterglow. This is particularly important in GRBs that have reported blackbody components at low energies in their XRT spectra. In V18 and V21, we have presented 19 GRBs that have significant blackbodies in their XRT spectra, and only two of these showed some signs of a decreasing $N_{\mathrm{H,intr}}$. 
In these GRBs, the blackbody interpretation is favoured in GRB~111225A, while we cannot exclude the scenario that the blackbody in GRB~171222A is an artefact of decreasing absorption.

\begin{figure*}[htb!]
\gridline{\fig{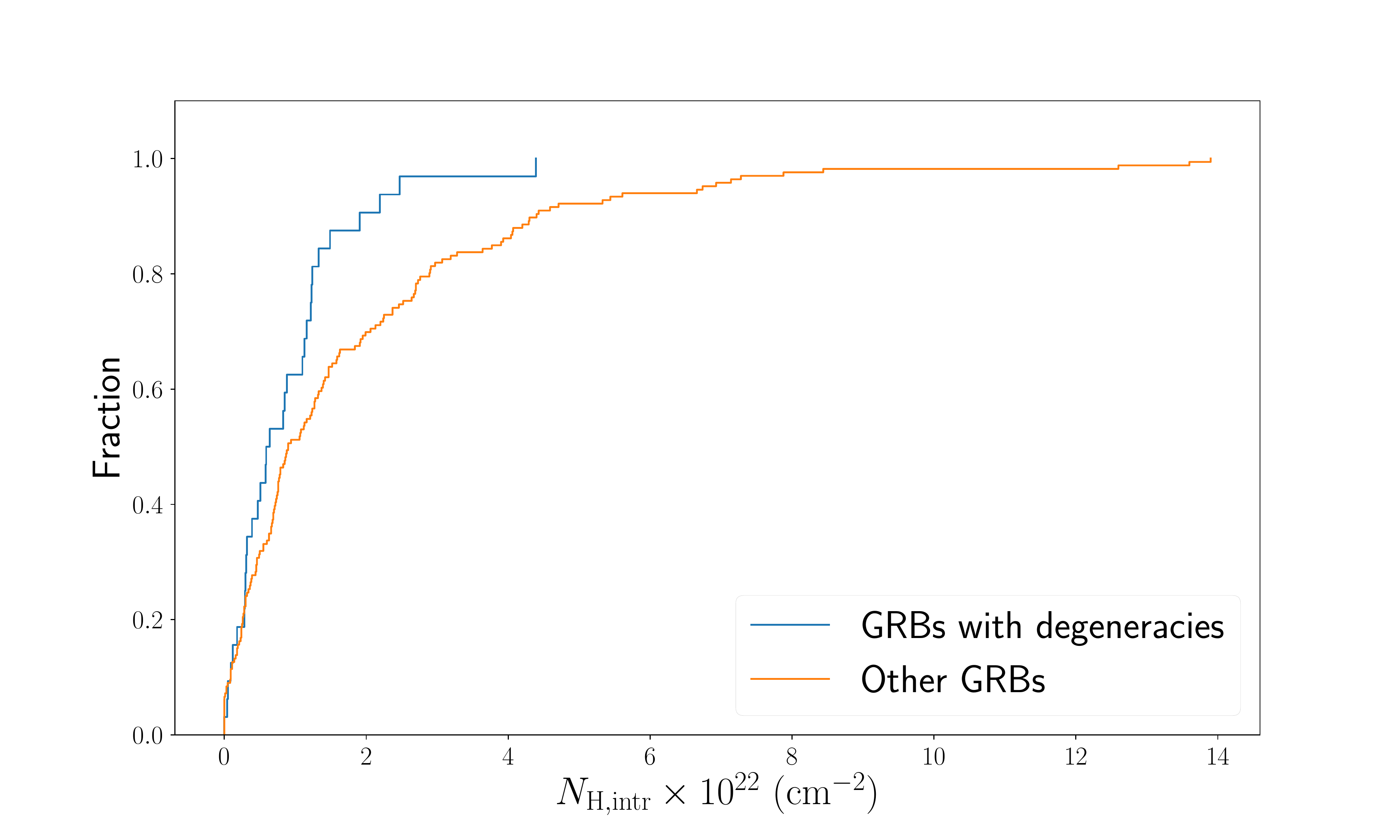}{0.49\textwidth}{}
          \fig{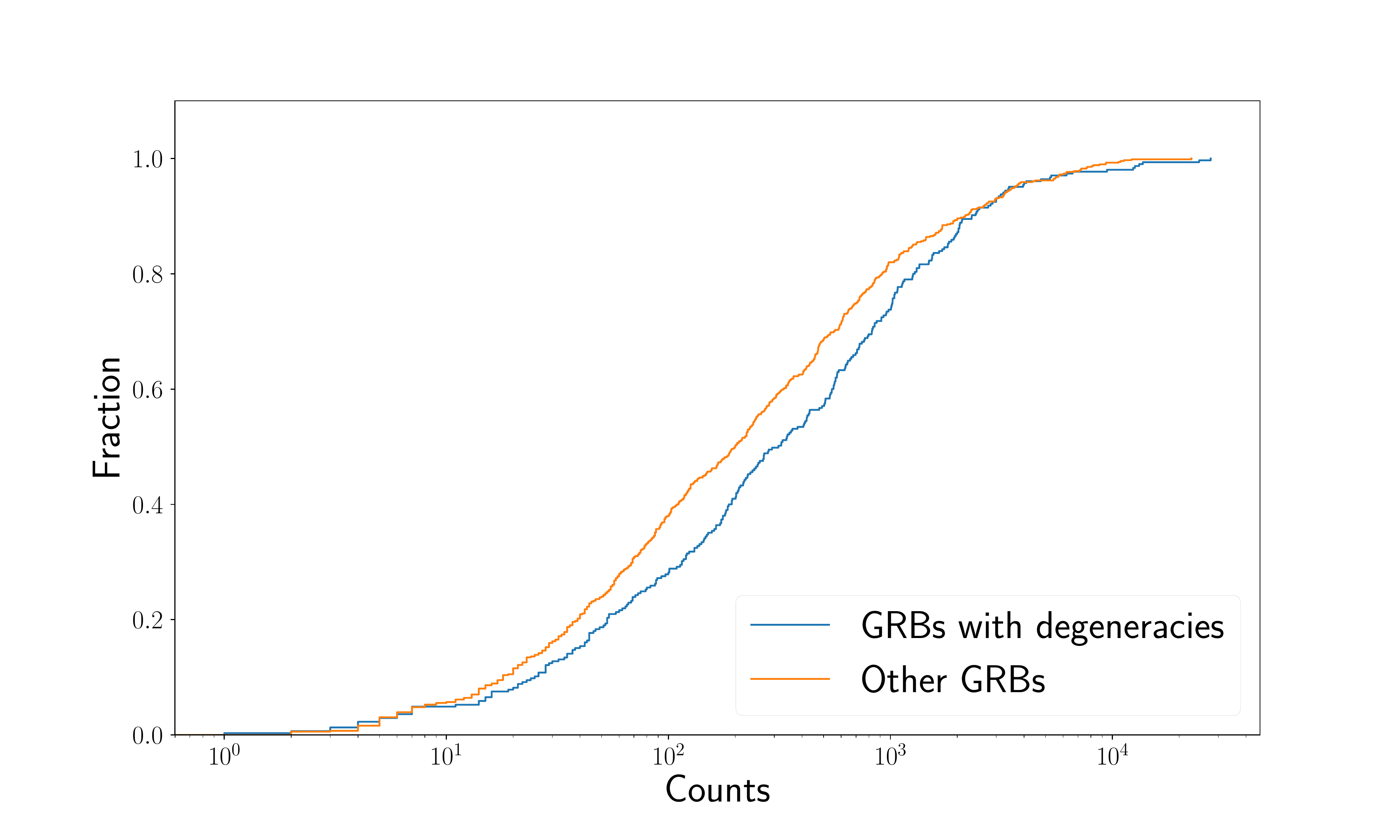}{0.49\textwidth}{}
          }
\gridline{\fig{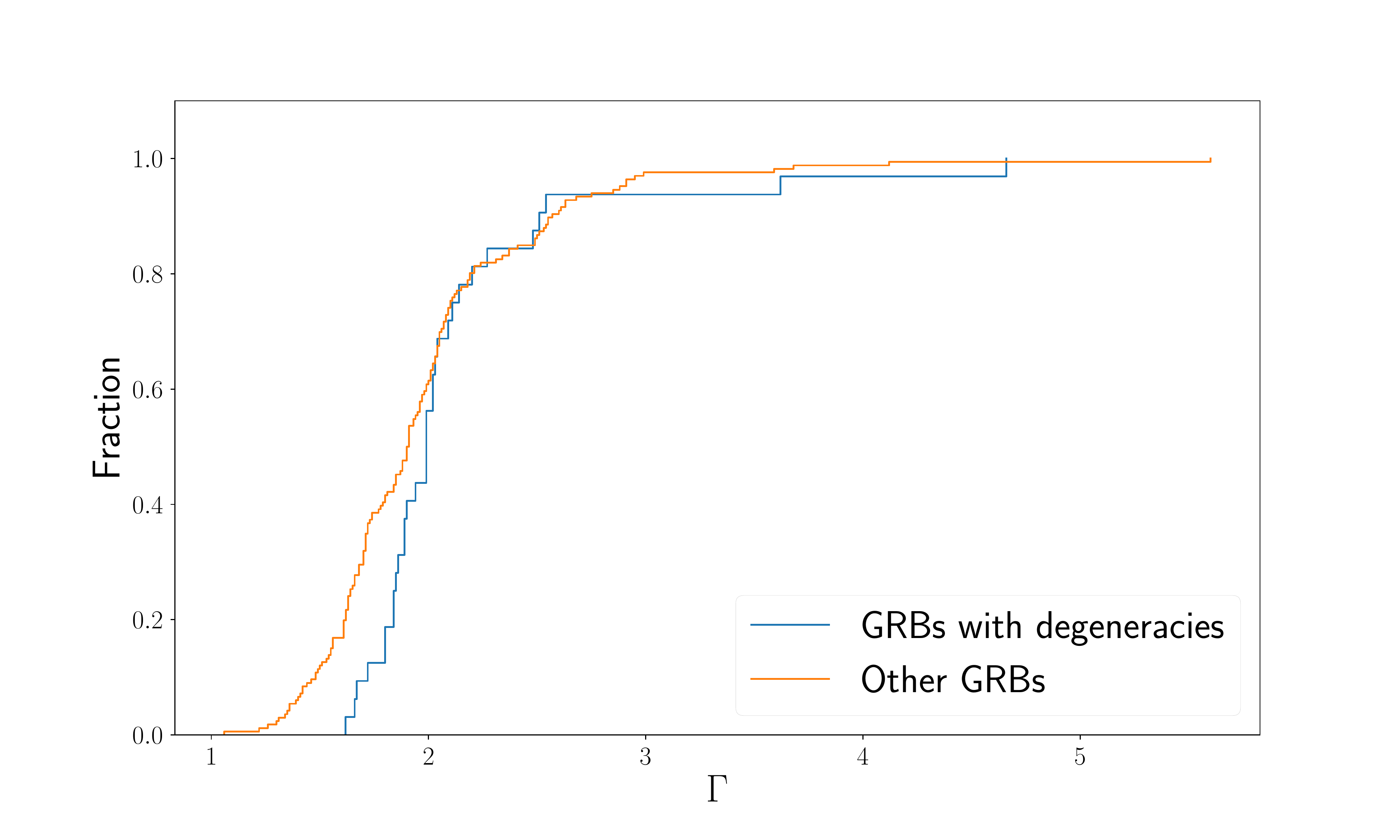}{0.49\textwidth}{}
          }
\caption{CDFs of $N_{\mathrm{H,intr}}$, total number of counts per spectrum and $\Gamma$ for GRBs with and without strong degeneracies between $N_{\mathrm{H,intr}}$ and $\Gamma$}.\label{fig:deghistograms}
\end{figure*}

It is clear that high-quality X-ray observations at earlier times than offered by {\it Swift}~XRT are needed to study the environments of GRBs through X-ray absorption. One of the challenges with determining absorption from current data is the spectral degeneracies between $N_{\mathrm{H,intr}}$ and $\Gamma$ (example plotted in Figure~\ref{fig:strongdegeneracies}). Inspection of all the corner posterior plots showed that there are 33 GRBs that exhibit similar or stronger degeneracies out of the whole sample of 199 GRBs. 
We used the Anderson-Darling (AD) test to investigate whether the GRBs that show strong degeneracies between $N_{\mathrm{H,intr}}$ and $\Gamma$ differ from the population as a whole in terms of any other properties. We found that these two groups have consistent distributions in redshift, $t_{\mathrm{90}}$ and observed flux, but that they do differ in total number of counts per spectrum, $\Gamma$ and $N_{\mathrm{H,intr}}$. 
The GRBs with strong degeneracies tend to have lower values of $N_{\mathrm{H,intr}}$, softer spectra with higher $\Gamma$, as well as a higher number of counts, as illustrated by the plots of the cumulative distribution functions (CDFs) in Figure~\ref{fig:deghistograms}. The AD test p-values are 0.001 for $N_{\mathrm{H,intr}}$ and $\Gamma$ (the lowest value that can be provided by SciPy's adaptation of the AD test that we used) and 0.003 for the counts. It is not surprising that spectra with low levels of absorption are more prone to degeneracies. Such spectra also tend to be softer and have higher count rates, as reflected by the differences in those distributions.  

Improved sensitivity with future telescopes will  improve these issues, while the possibility to resolve individual absorption lines will enable a much more detailed characterisation of the absorbing medium. 

\section{Summary and Conclusions} \label{sec:conclusions}

In this paper we have presented a systematic time-resolved spectral analysis of the X-ray spectra of 199 GRBs observed by \textit{Swift}~XRT between 2005 Apr 01 and 2018 Dec 31. We fitted the spectra with an absorbed power-law model with $N_{\mathrm{H,intr}}$ allowed to vary in order to search for signs of $N_{\mathrm{H,intr}}$ decreasing with time. Such a decrease would be a clear sign that the GRB is ionizing the matter in its vicinity and give information about the imminent environment of the GRB. We have structured the analysis as a Bayesian inference and used the posterior corner plots to investigate possible degeneracies between fit parameters. We also explored alternative spectral models for GRBs that showed evidence for a decrease in $N_{\mathrm{H,intr}}$. 

The analysis reveals seven GRBs that show signs of decreasing $N_{\mathrm{H,intr}}$, with a decrease from $(0.11 - 5) \times 10^{22}\  \mathrm{cm^{-2}}$ at $\sim 50$~s after trigger  to $(0.02 - 3.7) \times 10^{22}\ \mathrm{cm^{-2}}$ over the following $\sim 50-150$~s. Decreasing absorption on these time scales is expected for ionization of dense pc-scale clouds and shells. 
However, a caveat is that alternative models for the spectral evolution cannot be fully ruled out. 
For the vast majority of the sample, no decrease of $N_{\mathrm{H,intr}}$ is  observed. This is consistent with the scenario that the absorption is dominated by larger scales of the host galaxies and/or the IGM. It also implies that any material located close to the GRBs is either not massive enough for a decrease in absorption to be detectable, or it is ionized before the observations start.

Our results confirm that the standard assumption of constant $N_{\mathrm{H,intr}}$ made when fitting time-resolved X-ray spectra of GRBs is valid in the majority of GRBs. 
In line with this, we suggest that it is preferable to determine the total $N_{\mathrm{H,intr}}$ by simultaneously fitting time-resolved spectra of WT data with $N_{\mathrm{H,intr}}$ tied, but other spectral parameters free to vary. The values of $N_{\mathrm{H,intr}}$ obtained from WT data in this way are comparable to those obtained by fits to time-averaged PC spectra, but with a tighter constraints. 

Time evolution of $N_{\mathrm{H,intr}}$ affects in a small way the conclusions made about blackbodies in the early X-ray spectra of GRBs. Only two out of 19 GRBs with blackbody components identified in V18 and V21 showed signs of a decreasing $N_{\mathrm{H,intr}}$. In GRB~111225A, the power-law model is preferred only in the first bin, possibly suggesting that 
the blackbody component is subdominant at this time. 
However, in GRB~171222A the situation is more complex, with the preference between models alternating with time, and we cannot exclude the possibility that the observed blackbody is an artefact of decreasing $N_{\mathrm{H,intr}}$.

Finally, we have investigated the effect of spectral degeneracies in the analysis. We find that if statistics are good and the absorption high, there is little or no spectral degeneracy between $N_{\mathrm{H,intr}}$ and $\Gamma$. All these results show that high-quality X-ray data at earlier times is needed in order to fully probe the GRB environment through decreasing absorption.

\begin{acknowledgments}
This research was supported by the Knut and Alice Wallenberg Foundation. This work made use of data supplied by the UK Swift Science Data Centre at the University of Leicester.
\end{acknowledgments}

%

\vspace{5mm}
\facilities{Swift(XRT)}


\software{XSPEC \citep{xspec},  
          pymultinest \citep{Buchner2014}
          }



\appendix
\section{Cutoff power law fits} \label{appendix:bayesfactor}

\begin{figure*}[htb!]
\gridline{\fig{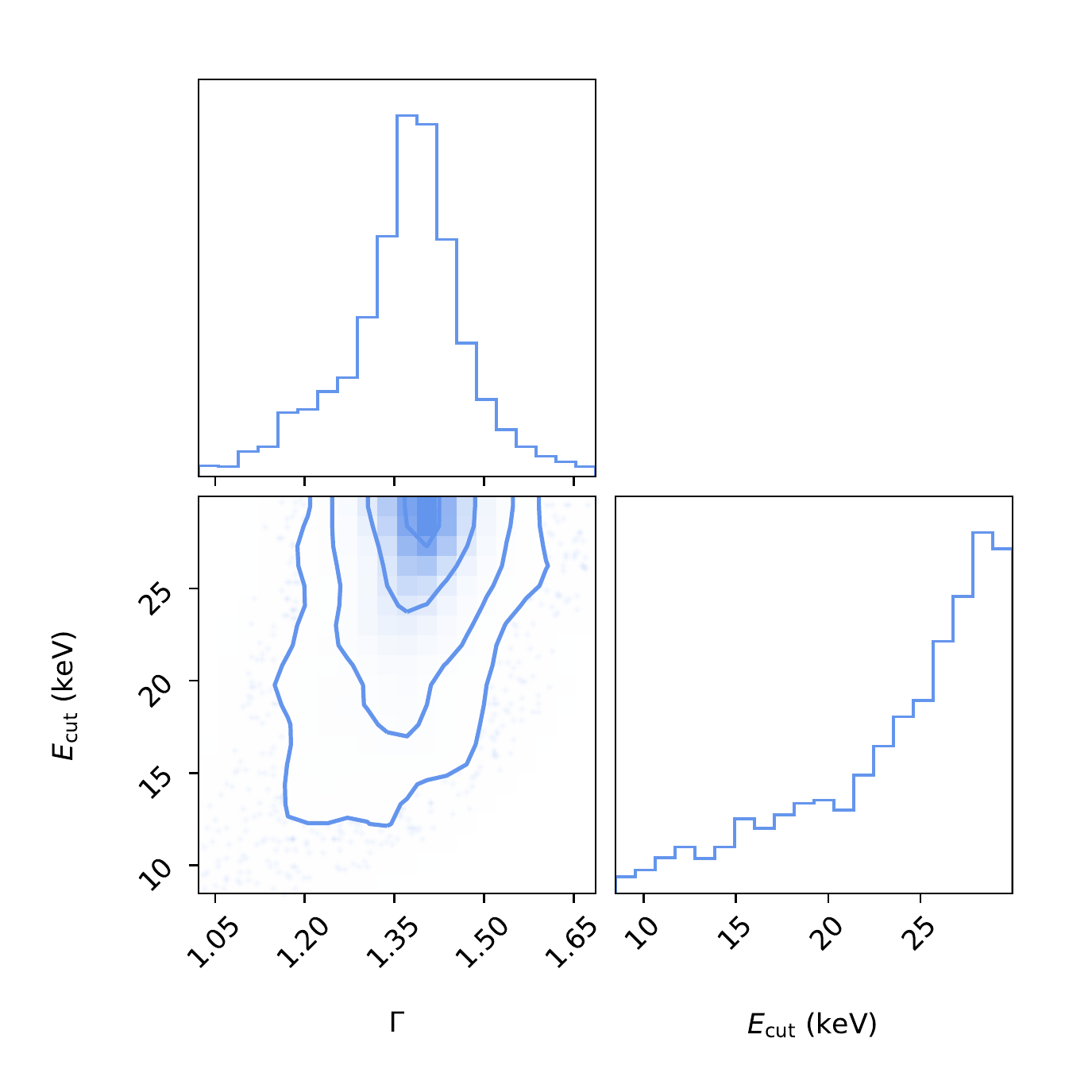}{0.49\textwidth}{}
            \fig{figures/GRB070306_bin1_cutoffpl.pdf}{0.5\textwidth}{}}
\gridline{\fig{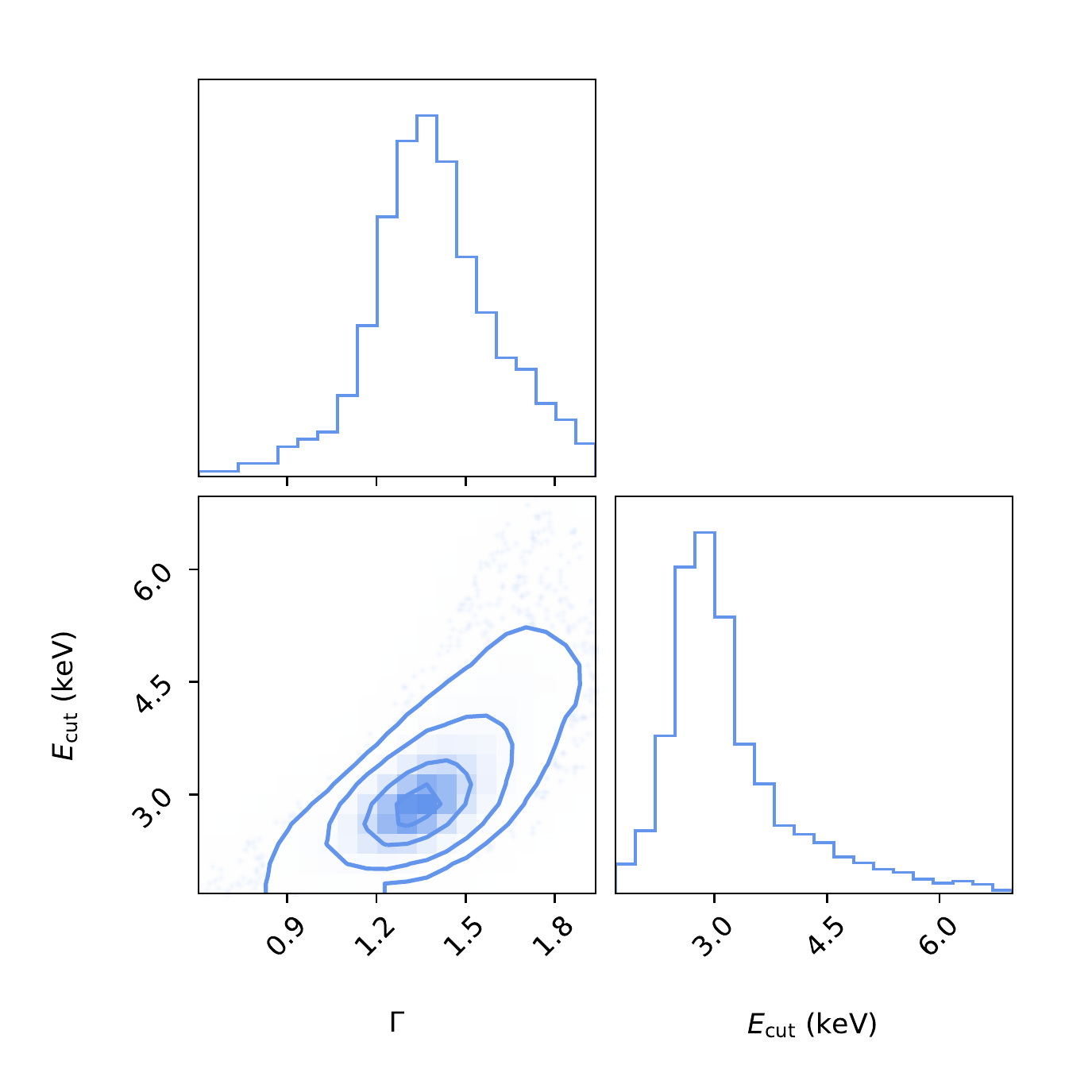}{0.49\textwidth}{}
    \fig{figures/gRB170607_bin3_cutoffpl.pdf}{0.5\textwidth}{}} 

\caption{Corner plots showing the posteriors for $E_{\mathrm{cut}}$ and $\Gamma$ (left) and the corresponding spectra with the best-fit model (right). GRB~070306 (top row, bin 1, time interval 67.1--72.5~s) illustrates the case where the AIC prefers power-law model with free $N_{\mathrm{H,intr}}$, while GRB~170607A (bottom row, bin 3,  time interval 79.2--87.1~s) represents the case where AIC prefers cutoff power-law model with fixed $N_{\mathrm{H,intr}}$.}
\label{fig:cutoffplfig}
\end{figure*}

Here we present results for the mean of the the marginalized posterior for $\Gamma$ and $E_{\mathrm{cut}}$ from the fits with the cutoff power-law model. The results for all bins of each burst are provided in Table \ref{tab:cutoffpl}. We also show two examples of posterior distributions alongside spectra with best-fit parameters in Figure~\ref{fig:cutoffplfig}, illustrating one case where the absorbed power-law model is preferred and one where the cutoff power-law model is preferred.

\begin{deluxetable}{cccc}[htb!]
\tablecaption{Mean of the marginalized posterior for $\Gamma$ and $E_{\mathrm{cut}}$ for the cutoff power-law model.
\label{tab:cutoffpl}}
\tablewidth{0pt}
\tablehead{
\colhead{GRB} & time bin\tablenotemark{a} & \colhead{$\Gamma$} & $E_{\mathrm{cut}}$\\
\colhead{} &\colhead{} &\colhead{} & \colhead{keV}} 
\startdata
070306 & bin 1 & $1.38 \pm 0.10$ & $25.53 \pm 5.20$ \\
070306 & bin 2 & $1.96 \pm 0.24$ & $8.12 \pm 6.20$ \\
070306 & bin 3 & $1.49 \pm 0.29$ & $1.69 \pm 0.24$ \\
070306 & bin 4 & $2.41 \pm 0.18$ & $22.49 \pm 6.50$ \\
070306 & bin 5 & $2.44 \pm 0.20$ & $22.23 \pm 7.10$ \\
070306 & bin 6 & $2.23 \pm 0.18$ & $21.20 \pm 7.50$ \\
070306 & bin 7 & $2.42 \pm 0.20$ & $20.38 \pm 7.50$ \\
\hline
071112C & bin 1 & $1.18 \pm 0.15$ & $2.22 \pm 0.27$ \\
071112C & bin 2 & $1.54 \pm 0.21$ & $6.18 \pm 4.72$ \\
071112C & bin 3 & $1.68 \pm 0.11$ & $23.17 \pm 6.20$ \\
071112C & bin 4 & $1.78 \pm 0.5$ & $22.17 \pm 7.20$ \\
\hline
081007 & bin 1 & $2.60 \pm 0.24$ & $19.79 \pm 7.71$ \\
081007 & bin 2 & $2.78 \pm 0.22$ & $15.90 \pm 8.10$ \\
081007 & bin 3 & $3.15 \pm 0.28$ & $21.12 \pm 7.50$ \\
\hline
090926B & bin 1 & $1.03 \pm 0.15$ & $2.23 \pm 0.25$ \\
090926B & bin 2 & $1.62 \pm 0.29$ & $3.81 \pm 2.00$ \\
090926B & bin 3 & $2.00 \pm 0.11$ & $22.42 \pm 5.84$ \\
090926B & bin 4 & $1.87 \pm 0.11$ & $22.60 \pm 7.50$ \\
090926B & bin 5 & $1.55 \pm 0.13$ & $22.07 \pm 7.50$ \\
\hline
170607A & bin 1 & $0.84 \pm 0.10$ & $0.97 \pm 0.03$ \\
170607A & bin 2 & $1.31 \pm 0.10$ & $3.01 \pm 0.35$ \\
170607A & bin 3 & $1.38 \pm 0.20$ & $2.98 \pm 0.76$ \\
170607A & bin 4 & $1.62 \pm 0.11$ & $4.95 \pm 1.05$ \\
170607A & bin 5 & $1.70 \pm 0.08$ & $2.77 \pm 0.24$ \\
170607A & bin 6 & $2.20 \pm 0.07$ & $26.30 \pm 4.63$ \\
170607A & bin 7 & $2.30 \pm 0.10$ & $26.11 \pm 4.77$ \\
170607A & bin 8 & $2.43 \pm 0.11$ & $25.43 \pm 5.24$ \\
170607A & bin 9 & $2.58 \pm 0.08$ & $24.40 \pm 5.50$ \\
170607A & bin 10 & $2.43 \pm 0.14$ & $24.73  \pm 5.76$ \\
\enddata
\tablenotetext{a}{Time bins correspond to the ones plotted in Figure \ref{fig:Nh_timeevolution}.}
\end{deluxetable}

\bibliography{sample631}{}
\bibliographystyle{aasjournal}



\end{document}